\documentclass[a4paper,11pt]{article}
\pdfoutput=1
\usepackage{jheppub,MnSymbol}
\usepackage{booktabs, hyperref}
\usepackage[math]{cellspace}
\newcommand{\beq}{\begin{equation}}
\newcommand{\eeq}{\end{equation}}
\newcommand{\bea}{\begin{eqnarray}}
\newcommand{\eea}{\end{eqnarray}}
\newcommand{\bfig}{\begin{figure}}
\newcommand{\efig}{\end{figure}}
\newcommand{\bc}{\begin{center}}
\newcommand{\ec}{\end{center}}
\newcommand{\dd}{\mathrm{d}}

\DeclareMathOperator{\Tr}{Tr}

\def\bom#1{{\mbox{\boldmath $#1$}}}
\def\to{\rightarrow}
\def\d{{\rm d}}
\def\eps{\epsilon}
\def\mgl{m_{\tilde g}}
\def\msq{m_{\tilde q}}
\def\msqc{m_{\tilde{q}_c}}
\def\mga{m_{\tilde \chi}}
\let\alignoriginal\align
\let\endalignoriginal\endalign
\renewenvironment{align}{\spreadlines{1em} \alignoriginal}{\endalignoriginal \endspreadlines}
\def\bsp#1\esp{\begin{split}#1\end{split}}

\preprint{MS-TP-16-09}

\title{Soft gluon resummation for associated gluino-gaugino production at the LHC}

\author[a,b]{Benjamin Fuks}
\author[c]{\!\!,\ Michael Klasen}
\author[c]{\!\! and Marcel Rothering}

\affiliation[a]{Sorbonne Universit\'es, UPMC Univ.~Paris 06, UMR 7589, LPTHE, F-75005 Paris, France}
\affiliation[b]{CNRS, UMR 7589, LPTHE, F-75005 Paris, France}
\affiliation[c]{Institut f\"ur Theoretische Physik, Westf\"alische Wilhelms-Universit\"at
 M\"unster, Wilhelm-Klemm-Stra\ss{}e 9, D-48149 M\"unster, Germany}

\emailAdd{fuks@lpthe.jussieu.fr}
\emailAdd{michael.klasen@uni-muenster.de}
\emailAdd{marcel.rothering@uni-muenster.de}

\abstract{We perform a threshold resummation calculation for the associated production of
gluinos and gauginos at the LHC to the next-to-leading logarithmic accuracy. Analytical
results are presented for the process-dependent soft anomalous dimension and the hard function.
The resummed results are matched to a full next-to-leading order calculation,
for which we have generalised the previously known results to the case of supersymmetric scenarios featuring non-universal squark masses.
Numerically, the next-to-leading logarithmic contributions increase the total
next-to-leading order cross section by 7 to 20\% for central scale choices and
gluino masses of 3 to 6 TeV, respectively,
and reduce its scale dependence typically from up to $\pm12$\% to
below $\pm3$\%.}


\begin{document}
\maketitle
\flushbottom

\section{Introduction}
\label{sec:1}

Supersymmetry (SUSY) is an attractive extension of the Standard Model (SM) of
particle physics. As the maximal space-time symmetry, it relates bosons to
fermions and predicts the existence of
spin partners of the SM particles that lead to a stabilisation
of the Higgs mass, to the unification of the gauge couplings at high energies,
and to a viable dark matter candidate. In many scenarios of the Minimal Supersymmetric
Standard Model
(MSSM), the dark matter candidate is the lightest neutralino, a mixed fermionic
state composed of the superpartners of the photon, the $Z$-boson and the $CP$-even
neutral Higgs bosons.

The search for SUSY particles is consequently an important research focus at the
LHC. Squarks and gluinos would be most copiously produced there through the strong
interaction, and their masses could therefore already be constrained by ATLAS and CMS in Run I
and the first year of Run II to lie beyond 1 TeV \cite{Aad:2016jxj,CMS:2015vqc}.
In contrast, pairs of sleptons and gauginos would be produced electroweakly. Their
mass limits are therefore still considerably lower, \textit{i.e.}~in the range of a few
hundreds of GeV \cite{Aad:2014vma,Khachatryan:2014qwa}. In most cases, the LHC analyses are based
on simplified scenarios with cascade decays of the squarks and gluinos to jets and
leptonic decays of the sleptons and gauginos, all accompanied by missing transverse
energy from the escaping lightest neutralino.

The experimental analyses rely on the availability of precise theoretical predictions
for the production cross sections. For many years, the state of the art were next-to-leading
order (NLO) calculations, which typically lead to an increase over the leading-order (LO) cross
section \cite{Dawson:1983fw,Baer:1990rq,Klasen:2006kb} and a reduction of the theoretical
uncertainty due to a stabilisation of the renormalisation and factorisation scale dependence
\cite{Beenakker:1996ed,Berger:1998kh,Beenakker:1999xh,Spira:2002rd,Berger:1999mc,%
Berger:2000iu,Berger:2002vd,GoncalvesNetto:2012yt,Binoth:2011xi,Goncalves:2014axa,Degrande:2014sta,Degrande:2015vaa}. Since SUSY particles
are often produced close to threshold, large logarithms can spoil the convergence of the
perturbative series. The current state state of the art is therefore to include also the
resummation of leading logarithms (LL) and next-to-leading logarithms (NLL),
that has been developed
originally for SM processes \cite{Sterman:1986aj,Kidonakis:1997gm,Kidonakis:1998bk,%
Catani:1989ne,Catani:1990rp}, in the context of slepton production~\cite{Bozzi:2004qq,Bozzi:2006fw,Bozzi:2007qr,%
Bozzi:2007tea,Fuks:2013lya}, gaugino production~\cite{Debove:2008nr,Debove:2009ia,Debove:2010kf,%
Debove:2011xj,Fuks:2012qx,Fuks:2013vua}, as well as squark and gluino production~\cite{Kulesza:2008jb,Langenfeld:2009eg,Falgari:2012hx,Langenfeld:2012ti,Beenakker:2009ha,%
Beenakker:2011sf,Beenakker:2013mva}. In some cases, also next-to-next-to-leading logarithms
(NNLL) and beyond have been resummed \cite{Beneke:2010da,Broggio:2011bd,Broggio:2013cia,Broggio:2013uba,Langenfeld:2012ti,Pfoh:2013edr,Beenakker:2011sf,Beenakker:2013mva,Beenakker:2014sma,Beenakker:2016gmf}, partly using soft-collinear effective theory (SCET). SCET and perturbative QCD results have been compared analytically and numerically, e.g., in Refs.\ \cite{Bonvini:2012az,Bonvini:2014qga}.
Since the full NNLO results and two-loop matching coefficients for gluino-gaugino associated
production are unknown, we for consistency present results only at the NLO+NLL level, even
though the two-loop soft anomalous dimensions are in principle known for arbitrary processes in
QCD, but are in practice not trivial to extract for our specific process. We therefore have to
leave a consistent NNLO+NNLL calculation for future work. Nevertheless, an estimate of approximate
NNLO+NNLL vs.\ NLO+NLL effects can be obtained from a recent calculation for stop pair production
\cite{Beenakker:2016gmf}, where they typically (for a stop mass of 2 TeV and an LHC energy of
13 TeV) amount to an increase of the total cross section by $5\%$ and a further stablisation
with respect to scale variations. Note that NLO calculations
have also been combined with parton showers for a variety of SUSY \cite{Jager:2012hd,Jager:2014aua,Gavin:2013kga,Gavin:2014yga,Baglio:2016rjx} and GUT processes \cite{Fuks:2007gk,Weydert:2009vr,Klasen:2012wq,Bonciani:2015hgv}, and these calculations usually agree well with resummation calculations within the theoretical uncertainties.

In this paper, we present a threshold resummation calculation for the associated production
of gluinos and gauginos at the NLL+NLO accuracy. This is one of two channels (the other one
being the associated production of squarks and gauginos, left for future work) for which NLO
calculations have
been computed previously \cite{Spira:2002rd,Berger:1999mc,Berger:2000iu,Berger:2002vd}, but
where a resummation calculation has so far not been performed. Its production cross section
is of intermediate strength, as it involves both strong and weak couplings. It can become
phenomenologically relevant in particular in the case that gluino pair production is
beyond the current LHC reach due to an exceedingly large gluino mass. This could very well be
realised in Nature, as in Grand Unified Theories (GUTs) one expects the gaugino mass parameters
$M_{i}$
to unify, similarly to the corresponding gauge couplings $\alpha_i$, so that after renormalisation
group running $M_3\simeq6M_1$ at the weak scale. The gluino with mass $\mgl=M_3$ is then typically
much heavier than the electroweak gauginos with masses of the order of the bino ($M_1$),
wino ($M_2\simeq2M_1$) or higgsino ($\mu$) mass parameters.

The outline of this paper is as follows: In Sec.~\ref{sec:2}, we describe briefly our
analytical calculations at LO and NLO and the refactorisation of the cross section, then
in more detail the necessary calculation of the hard
matching coefficient as well as the matching procedure and inverse Mellin transform.
Our numerical results are presented in Sec.~\ref{sec:3} for a typical benchmark scenario
that satisfies Higgs mass, flavour-changing neutral current, muon magnetic moment and LHC
constraints. Our conclusions are given in Sec.~\ref{sec:4}. The coupling conventions
employed in this paper are listed in App.~\ref{sec:a}, and the calculation of the soft
anomalous dimension is presented in App.~\ref{sec:b}.

\section{Soft gluon resummation}
\label{sec:2}

We start the presentation of our work with a description of our analytical results.
Our LO and NLO calculations are presented in Sec.~\ref{sec:2.1} and verified to
agree with those obtained previously in the limit of degenerate squark masses.
Sec.~\ref{sec:2.2} describes briefly the
refactorisation and resummation formalism up to the NLL accuracy. In Sec.~\ref{sec:2.4} and
App.\ \ref{sec:b}, we present in more detail the required calculations of the
process-dependent hard matching coefficient and the soft anomalous dimension.
Both are analytically checked against each other in Sec.~\ref{sec:2.5}, where
also the matching to the NLO calculation and the inverse Mellin transform are performed.

\subsection{Production of gluinos and gauginos at leading and next-to-leading order}
\label{sec:2.1}

At hadron colliders, the associated production of a gluino and a gaugino with four-momenta
$p_1$ and $p_2$ and masses $\mgl$ and $\mga$ proceeds at leading order (LO) through the annihilation
of a quark and an antiquark, both taken as massless, with four-momenta $p_a$ and $p_b$,
\beq
 q(p_a)\bar{q}'(p_b)\to\tilde{g}(p_1)\tilde{\chi}^{0,\pm}_j(p_2).
\eeq
Here, we distinguish possibly different quark flavours with a prime and label the gaugino mass
eigenstate by the index $j$ ($j=1,...,4$ for neutralinos and $j=1,2$ for charginos).

\begin{figure}
 \centering
 \includegraphics[width = 0.47\textwidth]{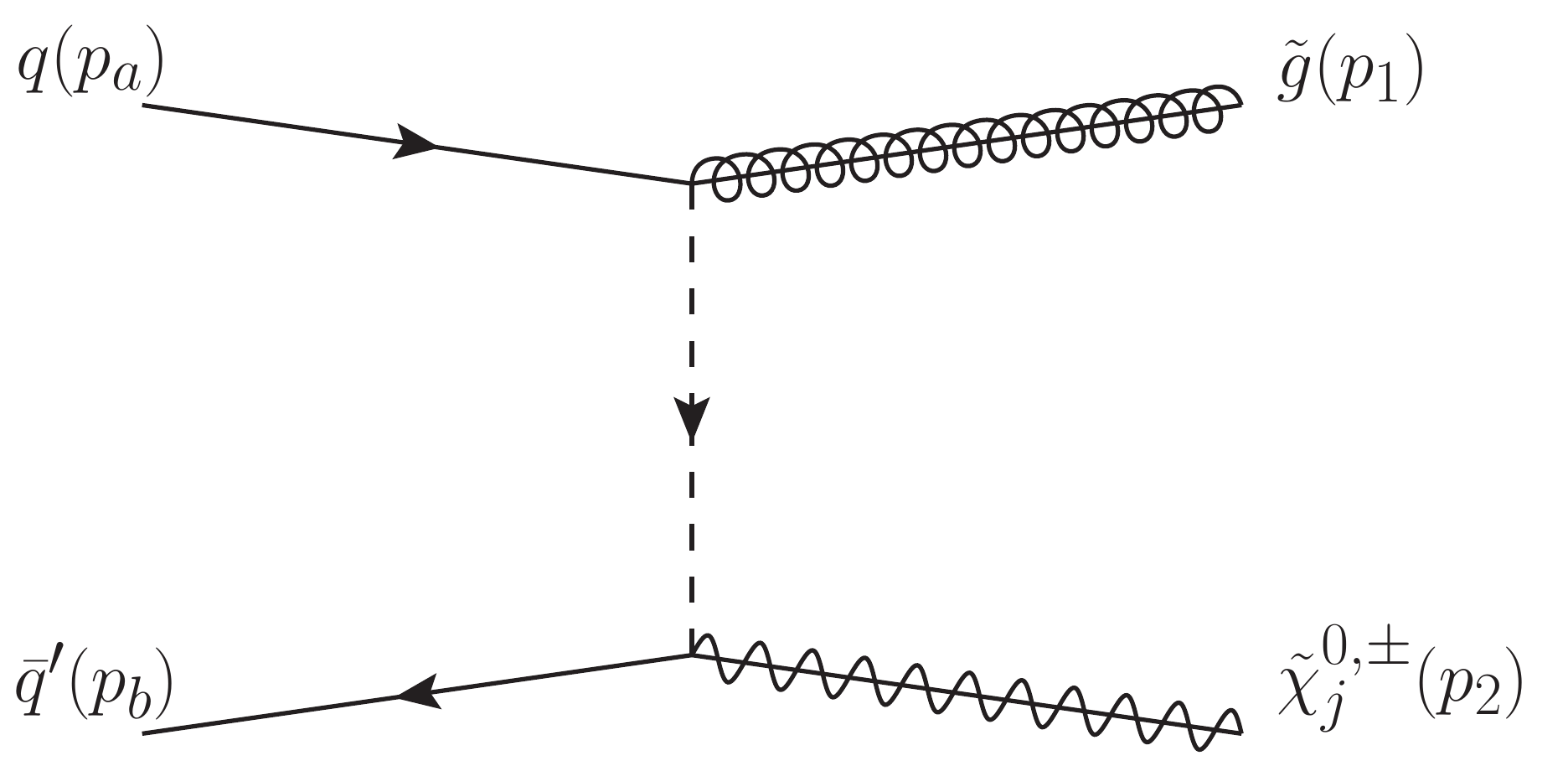}\hfill
 \includegraphics[width = 0.50\textwidth]{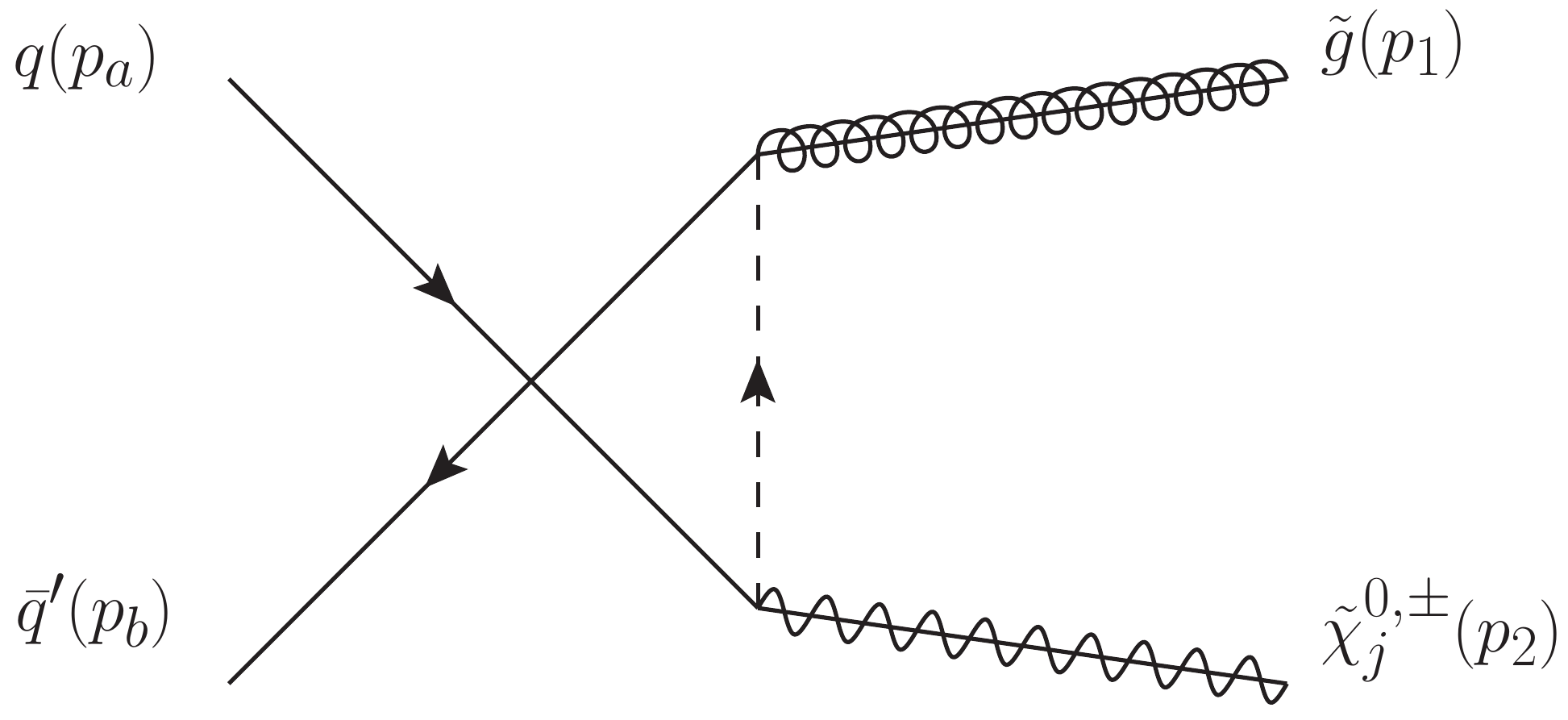}
 \caption{Tree-level $t$- (left) and $u$-channel (right) Feynman diagrams for the associated
 production of a gluino and a gaugino at hadron colliders. The dashed lines represent squark
 exchanges.}
 \label{fig:01} 
\end{figure}

The contributing processes are shown in Fig.~\ref{fig:01}. They are mediated by the exchange
of a virtual squark $(\tilde{q})$ in the $t$- (left) and $u$-channel (right), where $s=(p_a+p_b)^2
=(p_1+p_2)^2$, $t=(p_a-p_1)^2=(p_b-p_2)^2$ and $u=(p_a-p_2)^2=(p_b-p_1)^2$ satisfying $s+t+u=\mgl^2+
\mga^2$ are the usual partonic Mandelstam variables. The corresponding squared matrix elements
are given by
\bea
 \mathcal{M}_t\mathcal{M}^*_{t_c} &=& \frac{C_A C_F\,e\,g_s(\mu_r)}{(\msq^2 - t) (\msqc^2 - t)} ( \mathcal{L}' \mathcal{L}'_c + \mathcal{R}' \mathcal{R}'_c) (L L_c + R R_c)(\mgl^2 - t) (\mga^2 - t),
\\
 \mathcal{M}_u\mathcal{M}^*_{u_c} &=& \frac{C_A C_F\,e\,g_s(\mu_r)}{(\msq^2 - u) (\msqc^2 - u)}( \mathcal{L} \mathcal{L}_c + \mathcal{R} \mathcal{R}_c) (L' L'_c + R' R'_c)(\mgl^2 - u) (\mga^2 - u),
\\
\mathcal{M}_t \mathcal{M}^*_{u_c} &=& \frac{C_A C_F\,e\,g_s(\mu_r)}{(\msq^2 - t)(\msqc^2 - u)} \biggl[ \left(-s^2 + t^2 + u^2 + (\mga^2 + \mgl^2 ) (s - t - u) + 2 \mgl^2 \mga^2\right) \nonumber \\
&\times& \left( L L_c \mathcal{L}' \mathcal{L}'_c + R R_c \mathcal{R}' \mathcal{R}'_c \right) 
+ 2 \mgl \mga s (\mathcal{R} R_c  \mathcal{L}' L'_c   + \mathcal{L} L_c   \mathcal{R}' R'_c) \biggr],
\eea
where the label $c$ of the squark masses $m_{\tilde{q}}$ and couplings $L^{(')}$, $R^{(')}$,
${\cal L}^{(')}$ and ${\cal R}^{(')}$ refers to those
appearing in the complex conjugate diagrams. Our conventions for the notations of
couplings can be found in App.~\ref{sec:a}. The SU(3) colour factors are $C_A=3$ and $C_F=4/3$.
The relation between the electromagnetic and weak couplings $e=g\sin\theta_W$ depends on the weak
mixing angle $\theta_W$, and $g_s(\mu_r)$ is the (renormalisation-scale dependent) strong
coupling constant. The total spin- and colour-averaged squared amplitude is then
\beq
 \overline{|{\cal M}|^2} = \frac{1}{4C_A^2} \sum_{\substack {\tilde{q},\tilde{q}_c}} \left(\mathcal{M}_t\mathcal{M}^*_{t_c} + \mathcal{M}_u\mathcal{M}^*_{u_c} - 2 \operatorname{Re}(\mathcal{M}_t \mathcal{M}_{u_c}^*) \right)\,,
\eeq
where the sum is performed over all squarks in the propagators and where we have inserted a
relative minus sign between the $t$- and $u$-channel for the crossing of a fermion line.
We have hence generalised the results of the literature~\cite{Dawson:1983fw,Baer:1990rq,%
Klasen:2006kb,Beenakker:1996ed,Berger:1998kh,Beenakker:1999xh,Spira:2002rd} by allowing for
arbitrary squark
mixings and mass eigenstates in the squark couplings and propagators, as we have already done
in our calculations for slepton \cite{Bozzi:2004qq,Bozzi:2006fw,Bozzi:2007qr,Bozzi:2007tea,%
Fuks:2013lya} and
gaugino pair production \cite{Debove:2008nr,Debove:2009ia,Debove:2010kf,%
Debove:2011xj,Fuks:2012qx,Fuks:2013vua}. This will in particular allow us to extend existing
studies of SUSY flavour violation \cite{Bozzi:2005sy,Bozzi:2007me,delAguila:2008iz,Fuks:2008ab,%
Fuks:2011dg,DeCausmaecker:2015yca}  to new processes. Integration over the two-particle phase space
dPS$^{(2)}= \dd t/(8 \pi s)$ leads to the total partonic cross section
\beq
 \sigma_{ab}(s)=\int \dd \sigma_{ab}(s) = \int \frac{1}{2s}\,
 \overline{|{\cal M|}^2}\, {\rm dPS}^{(2)}\,,
\eeq
and convolution with the factorisation-scale dependent parton distribution functions (PDFs)
$f_{a/A}(x_a,\mu_f)$ and $f_{b,B}(x_b,\mu_f)$
\bea
 \sigma_{AB}~=~\int M^2\frac{\dd\sigma_{AB}}{\dd M^2}(\tau)&=&\sum_{a,b} \int_0^1 \!
 \dd x_a \,\dd x_b \,\dd z[x_a f_{a/A}(x_a,\mu_f^2)] [x_b f_{b/B}(x_b,\mu_f^2)] \nonumber\\
 &\times& \,[z\,\dd\sigma_{ab}(z,M^2,\mu_r^2,\mu_f^2)]\,\delta(\tau-x_ax_bz) 
 \label{eq:HadFacX}
\eea
to the total hadronic cross section \cite{Collins:1989gx}. Here, $\tau=M^2/S$ denotes the ratio
of the squared invariant mass of the produced SUSY particle pair and the hadronic centre-of-mass
energy and is related to the partonic momentum fractions $x_a$ and $x_b$ by $z=\tau/(x_a x_b)$ with $z=1$
at LO.

For non-mixing squark exchanges, the NLO corrections are well-known \cite{Berger:1999mc,%
Berger:2000iu,Berger:2002vd,Spira:2002rd,GoncalvesNetto:2012yt}. They involve one-loop self-energy, vertex correction
and box diagrams interfering with those at tree level as well as squared real gluon and quark
emission
diagrams, which can involve intermediate on-shell squarks. We have recalculated the full NLO
cross section for general squark mixings and mass eigenstates using dimensional regularisation
of ultraviolet and infrared divergencies, $\overline{\rm MS}$ renormalisation for couplings and
wave functions substituted with a finite shift in the quark-squark-gluino Yukawa coupling to
restore SUSY \cite{Martin:1993yx}, on-shell renormalisation for all squark and gluino masses,
and the dipole subtraction method to deal with infrared and collinear divergences
\cite{Catani:1996jh,Catani:2002hc}. This method exploits the fact that the NLO cross section
$\sigma^{(1)}$
can be split into two parts $\sigma^{\{2\}}$ and $\sigma^{\{3\}}$ with two- and three-particle
kinematics, respectively, and a collinear counterterm $\sigma^C$, that removes initial-state
collinear singularities,
\beq
 \label{eq:2.8}
 \sigma^{(1)}=\sigma^{\{3\}}+\sigma^{\{2\}}+\sigma^C=
 \int_3[\dd\sigma^{R}-\dd\sigma^A]_{\eps=0}+\int_2[\dd\sigma^V+\int_1\dd\sigma^A]_{\eps=0}
 +\sigma^C.
\eeq
The two- and three-particle cross sections are individually regularised by subtracting from
the real corrections $\sigma^{R}$ an auxiliary cross section $\sigma^{A}$. The latter
captures all
infrared-singular behaviour, can be analytically integrated over the singular phase space
regions, and is added back to the virtual corrections $\sigma^A$. In the limit of
mass-degenerate squarks, we achieve full numerical agreement with our previous
calculation \cite{Berger:1999mc,Berger:2000iu,Berger:2002vd}, that employed the phase-space
slicing method, and the public code {\sc Prospino}~\cite{Beenakker:1996ed}.

\subsection{Refactorisation}
\label{sec:2.2}

Beyond LO, $z = M^2/s$ describes the energy fraction going into the hard scattering process.
The energy fraction carried by an additional emitted gluon (or quark) is then $1-z$ and
describes the distance from partonic threshold. As is well known, large logarithms
\beq
 \left({\alpha_s\over2\pi}\right)^n \left[{\ln^m(1-z)\over1-z}\right]_+
\eeq
with $m \leq 2n-1$ remain at higher orders in $\alpha_s=g_s^2/(4\pi)$ even after the cancellation
of soft and collinear divergences among the virtual and real emission corrections
\cite{Kinoshita:1962ur,Lee:1964is}. They arise due to restrictions on the phase space boundary
of the latter and spoil the convergence of the perturbative series when $z\rightarrow 1$,
\textit{i.e.}~for soft emitted gluons. They must therefore be resummed to all orders
in $\alpha_s$ for reliable predictions in the threshold region.

Resummation of these logarithmic corrections to all orders and exponentiation can be achieved
when the kinematical and dynamical parts of the cross section are fully factorised. By applying
a Mellin transform
\beq
 F(N)=\int_0^1 \dd y \,y^{N-1} F(y)\,,
  \label{eq:MelDef}
\eeq
to the quantities $F=\sigma_{AB}$, $\sigma_{ab}$, $f_{a/A}$ and $f_{b/B}$ with $y=\tau$, $z$, $x_a$
and $x_b$ in
Eq.~\eqref{eq:HadFacX}, the hadronic cross section can be written as a simple product
\beq
 M^2\frac{\dd\sigma_{AB}}{\dd M^2}(N-1) = \sum_{a,b} f_{a/A}(N,\mu_f^2) 
 f_{b/B}(N,\mu_f^2) \sigma_{ab}(N,M^2,\mu_f^2,\mu_r^2)\,.
 \label{eq:HadFacN}
\eeq
Large logarithms in $z\to1$ then turn into large logarithms of the Mellin variable $N$.

Applying the eikonal Feynman rules, using renormalisation group equation properties and the
evolution equations, the partonic cross section can then be refactorised and written in the
resummed form \cite{Sterman:1986aj,Kidonakis:1997gm,Kidonakis:1998bk}
\beq
 \label{eq:2.12}\bsp
 \sigma^{\rm (res.)}_{ab\to ij}(N,M^2,\mu^2) =&\ \sum_{I}\,\mathcal{H}_{ab\to ij,I}(M^2,\mu^2)\,
 \Delta_a (N,M^2,\mu^2)\,\Delta_b (N,M^2,\mu^2)\\ &\qquad \times \Delta_{ab\to ij,I}(N,M^2,\mu^2)\,,
\esp\eeq
where we have identified the renormalisation and factorisation scales $\mu=\mu_r=\mu_f$
for brevity. The hard function 
\beq
 \label{eq:2.13}
 \mathcal{H}_{ab\to ij,I}(M^2,\mu^2)=\sum_{n=0}^\infty a_s^n(\mu^2)
 {\cal H}^{(n)}_{ab\to ij,I}(M^2,\mu^2),
\eeq
which is non-singular when $z\to1$ or $N\to\infty$ can be expanded perturbatively in
\mbox{$a_s=\alpha_s/(2\pi)$} and is discussed in more detail in Sec.~\ref{sec:2.4}. Like the
soft wide-angle function $\Delta_{ab\to ij,I}$ to be discussed in App.~\ref{sec:b}, it is
sensitive to the colour structure of the underlying hard process from which the gluon has been
emitted, and can be decomposed into irreducible colour representations $I$.
Together with the functions $\Delta_{a,b}$ describing soft collinear radiation, the
soft wide-angle function $\Delta_{ab\to ij,I}$ can be exponentiated in a closed form
\cite{Vogt:2000ci}
\beq
 \label{eq:NNLL-expa}
 \Delta_a\Delta_b\Delta_{ab\to ij,I} = \exp\Big[L G^{(1)}_{ab}(\lambda)
 + G^{(2)}_{ab\to ij,I}(\lambda,M^2/\mu^2) 
 + \ldots \Big]  \,,
\eeq
with $\lambda = a_s \beta_0 L$, $L = \ln{\bar N}$ and ${\bar N}=Ne^{\gamma_E}$, which contains
all the enhanced logarithmic terms. The terms $G^{(1)}_{ab}$ and $G^{(2)}_{ab \rightarrow ij}$ represent
the leading-logarithmic (LL) and next-to-leading logarithmic (NLL) approximations
\cite{Sterman:1986aj,Kidonakis:1997gm,Kidonakis:1998bk,Catani:1989ne,Catani:1990rp}
\bea
G^{(1)}_{ab}(\lambda) &=& g_a^{(1)}(\lambda) + g_b^{(1)}(\lambda), \\
G^{(2)}_{ab \rightarrow ij}(\lambda) &=& g_a^{(2)}(\lambda,M^2,\mu_r^2, \mu_f^2) + g_b^{(2)}(\lambda,M^2,\mu_r^2, \mu_f^2) + h^{(2)}_{ab \rightarrow ij,I}(\lambda)\,, 
\eea
with
\bea
g_a^{(1)}(\lambda) &=& \frac{A_a^{(1)}}{2 \beta_0 \lambda} \left[ 2 \lambda + (1 - 2 \lambda) \ln{(1 - 2 \lambda)} \right]\,, \\
g_a^{(2)}(\lambda) &=& \frac{A_a^{(1)} \beta_1}{2 \beta_0^3} \left[ 2 \lambda + \ln{(1 - 2 \lambda) + \frac{1}{2} \ln^2{(1 - 2 \lambda)}} \right]\,, \nonumber\\
&-& \frac{A_a^{(2)}}{2 \beta_0^2} \left [2 \lambda + \ln{(1 - 2 \lambda)} \right] \nonumber\\
&+& \frac{A_a^{(1)}}{2 \beta_0} \left[\ln{(1 - 2 \lambda)} \ln{\left(\frac{M^2}{\mu_r^2} \right)} + 2 \lambda \ln{\left(\frac{\mu_f^2}{\mu_r^2} \right)} \right]\,,\\
h^{(2)}_{ab \rightarrow ij,I}(\lambda) &=& \frac{\ln{\left(1 - 2 \lambda\right)}}{2\beta_0}D^{(1)}_{ab \rightarrow ij,I}\,,
\label{resum}
\eea
where the one-loop coefficient $D^{(1)}_{ab \rightarrow ij,I}$ depends on the soft anomalous dimension
and is process dependent. While it vanishes for gaugino pair production with coloured
particles in the initial state only \cite{Debove:2010kf}, it must be included for associated
gluino-gaugino production, where soft gluons can also be emitted from the final-state gluino.
The associated collinear divergence is, however, screened due to the massive emitter. The
coefficients in the above equations read
\bea
\label{eq:2.20}
A^{(1)}_{a} &=& 2 C_a\, ,\\
A^{(2)}_{a} &=& 2 C_a \left[ \left( \frac{67}{18} - \frac{\pi^2}{6} \right) C_A - \frac{5}{9} n_f \right] \, ,\\
D^{(1)}_{ab \rightarrow ij,I} &=& \frac{2\pi}{\alpha_s} \operatorname{Re} (\bar{\Gamma}_{ab \rightarrow ij,II})\,, \label{eq:OL}
\eea
where $C_a = C_{F,A}$ for quarks and gluons and $\bar{\Gamma}_{ab \rightarrow ij,II}$ is the
modified diagonal soft anomalous dimension of the colour representation $I$.

\subsection{Hard matching coefficient}
\label{sec:2.4}

The resummation of logarithmically enhanced contributions at threshold can
be improved by including in the hard function ${\cal H}_{ab\to ij,I}(M^2,\mu^2)$
in Eq.\ (\ref{eq:2.13}) not only the LO cross section
\beq
 {\cal H}^{(0)}_{ab\to ij}(M^2,\mu^2)=\sigma^{(0)}_{ab\to ij}(M^2),
\eeq
but also the $N$-independent contributions of the NLO cross section
\beq
 {\cal H}^{(1)}_{ab\to ij}(M^2,\mu^2)=\sigma^{(0)}_{ab\to ij}(M^2)\,
 C^{(1)}_{ab\to ij}(M^2,\mu^2),
\eeq
which beyond NLO are multiplied by threshold logarithms. The hard matching coefficient
$C^{(1)}_{ab\to ij}(M^2,\mu^2)$ is obtained by computing the Mellin transform of the full
NLO corrections $\sigma^{(1)}_{ab\to ij}(M^2,\mu^2)/\sigma^{(0)}_{ab\to ij}(M^2)$ described
in Sec.\ \ref{sec:2.1}, keeping only the $N$-independent terms. In the case of
associated gluino-gaugino production, this is again simplified by the fact that there
is only one colour basis tensor, so that the index $I$ can be dropped.

In Eq.\ (\ref{eq:2.8}), the three-particle contributions to $\sigma^{(1)}$  can safely
be neglected, since all infrared divergences are canceled after subtracting from
$\dd\sigma^R$ the auxiliary cross section $\dd\sigma^A$ and the finite terms are
phase-space suppressed near threshold \cite{Beenakker:2011sf,Beenakker:2013mva}.
The virtual corrections $\dd\sigma^V$ and the integrated dipoles $\int_1\dd\sigma^A$ are
straightforward to transform into Mellin space in the invariant-mass threshold approach,
since they are proportional to \mbox{$\delta(1-z)$} and thus constant in $N$. The
corresponding analytical results can be found in Refs.\ \cite{Berger:2000iu} and
\cite{Catani:1996jh,Catani:2002hc}, respectively. The collinear
remainder $\sigma^C$ is usually split into contributions from two insertion operators
$\bom{P}$ and $\bom{K}$ \cite{Catani:2002hc}
\beq
 \sigma^C=\sum_{a'}\int_0^1\dd x\int_2
 [\dd\sigma^{(0)}_{a'b}(xp_a,p_b)\otimes\langle a'|\bom{P}+\bom{K}|a\rangle(x)
 + (a\leftrightarrow b)
 ]_{\eps=0}.
\eeq
The former is directly related to the regularised Altarelli-Parisi splitting distributions
at ${\cal O}(\alpha_s)$, while the latter depends on the factorisation scheme and on the
regular parts of the Altarelli-Parisi splitting distributions.
For the initial quark $a$, we obtain after transforming to Mellin space
\beq
\label{eq:2.44}
 \langle \bom{P}(N) \rangle = \frac{\alpha_s}{2 \pi}
 \biggl[ (2C_F-C_A)\ln\frac{\mu_f^2}{s}
              +C_A\ln\frac{\mu_f^2}{\mgl^2-t}\biggl]
 \biggl[\ln {\bar N} -{3\over4}\biggl] \,+\, \mathcal{O}\left(\frac{1}{N}\right)
\eeq
and
\bea
\label{eq:2.45}
 \langle \bom{K}(N)\rangle &=& \frac{\alpha_s}{2 \pi}
         \Bigg\{ 2 C_F \ln^2{\bar N}
               + C_A \left[\ln{\mgl^2\over \mgl^2-t} + 1\right] \ln {\bar N}
 +{\pi^2\over2} C_F-\gamma_q-K_q\\
 &+&{C_A\over4}\biggl[
 1+4\,{\rm Li}_2{2\mgl^2-t\over\mgl^2}
 +\bigg(1+4\ln{\mgl^2\over\mgl^2-t}
 +2{\mgl^2\over\mgl^2-t}\bigg)\ln{\mgl^2\over2\mgl^2-t} \nonumber\\
 & & \  +3\ln\bigg(1+{2\mgl\over\mgl^2-t}\left(\mgl-\sqrt{2\mgl^2-t}\ \right)\bigg)
     +6{\mgl\over\mgl+\sqrt{2\mgl^2-t}}-3 \biggl]
 \Bigg\} + \mathcal{O}\left(\frac{1}{N}\right)\nonumber
\eea
with 
\bea
 \gamma_q~=~ {3\over2}\,C_F  &\ ,\ & K_q ~=~ \left({7\over2}-{\pi^2\over6}\right)\,C_F\,,
\eea
and similarly for the incoming antiquark $b$ with $t\to u$. Non-diagonal operators
give only $1/N$-suppressed contributions, and there are no initial-state gluons at LO.
In the limit of $C_A\to0$, one recovers the well-known results for Drell-Yan like processes
\cite{Debove:2010kf}.

For the hard matching coefficient, only the $N$-independent parts of the above results
are needed. The logarithmic terms terms can be used to check the resummed cross
section when re-expanded to NLO (see below).
Since the $1/N$ terms have been systematically neglected, the collinear improvement
of the resummation formalism suggested in Refs.~\cite{Kramer:1996iq,Catani:2001ic}
has not been performed in contrast to our calculations for uncoloured slepton~\cite{Bozzi:2007qr}
and gaugino pair production~\cite{Debove:2010kf}, but similarly
to the calculations for coloured squarks and
gluinos~\cite{Beenakker:2009ha,Beenakker:2011sf,Beenakker:2013mva}.
This is in particular due to the fact that analytic results for
the subleading terms in the Mellin transform of the collinear remainder can not be obtained.
For the Drell-Yan process, the constant terms in the hard function ${\cal H}_{ab\to ij}^{(1)}
(M^2,\mu^2)$ are sometimes also exponentiated based on the argument that they factorise the
complete Born cross section, include finite remainders of the infrared singularities in the
virtual corrections and are thus related to the corresponding singularities in the real
corrections giving rise to the large logarithms \cite{Eynck:2003fn}. However, similarly to
gaugino pair production \cite{Debove:2010kf}, gluino-gaugino
associated production does not proceed through a single $s$-channel diagram (see
Fig.~\ref{fig:01}) and the virtual corrections factorise only at the level of amplitudes,
and not at the level of
the full cross section~\cite{Berger:1999mc,Berger:2000iu,Berger:2002vd}. Resumming these
finite terms is therefore not justified in this case.

\subsection{Matching and inverse Mellin transform}
\label{sec:2.5}

While near threshold the resummed cross section is a valid approximation,
far from it the normal perturbative calculation should be used. A reliable
prediction in all kinematic regions is then obtained through a consistent
matching of the two results with
\beq
 \sigma_{ab} = \sigma_{ab}^{\rm (res.)} + \sigma_{ab}^{\rm (f.o.)} - \sigma_{ab}^{\rm (exp.)}\,.
\eeq
Here, the resummed cross section $\sigma_{ab}^{\rm (res.)}$ in Eq.~\eqref{eq:2.12}
has been re-expanded to NLO, yielding $\sigma_{ab}^{\rm (exp.)}$, and subtracted from
the fixed-order calculation $\sigma_{ab}^{\rm (f.o.)}$ in Eq.~\eqref{eq:2.8} in order to avoid
the
double counting of the logarithmically enhanced contributions. At ${\cal O}(\alpha_s)$, we then
obtain
\bea
 \label{eq:2.48}
 \sigma_{ab}^{\rm (exp.)} &=&
 \mathcal{H}^{(0)}_{ab\to ij,I}(M^2,\mu^2)
 + \frac{\alpha_s}{2 \pi} \mathcal{H}^{(1)}_{ab\to ij,I}(M^2,\mu^2)
 + \frac{\alpha_s}{2 \pi} \, \mathcal{H}^{(0)}_{ab\to ij,I}(M^2,\mu^2)\nonumber\\
 &\times& \left[ (A_a^{(1)} + A_b^{(1)}) \ln^2 {\bar N} + 
 \left( (A_a^{(1)} + A_b^{(1)}) \ln{\frac{\mu_f^2}{s}}
 -2D^{(1)}_{ab \rightarrow ij,I} \right)\ln {\bar N} \right]\,, 
\eea
where $\mathcal{H}^{(0)}$ and $\mathcal{H}^{(1)}$ are the first and second order parts
of the hard matching coefficient (see Sec.~\ref{sec:2.4}).

Inserting $A_a^{(1)}=A_b^{(1)}=2C_F$ ({\it cf.}~Eq.~\eqref{eq:2.20}), the leading logarithms
$\alpha_s/(2\pi)\ln^2{\bar N}$ have the coefficient $4C_F$, which agrees with the leading
logarithmic contribution to the hard matching coefficient arising exclusively from the
$\bom{K}$-operators in the collinear remainder in Eq~\eqref{eq:2.45}. The coefficient $4C_F$
also governs the scale- and more precisely
the $\ln(\mu_f^2/s)$-dependent part of the next-to-leading logarithms $\alpha_s/(2\pi)\ln{\bar N}$,
which agrees with the corresponding parts of the quark and antiquark $\bom{P}$-operator expectation
values in the collinear remainder in Eq.~\eqref{eq:2.44}. In contrast, the $C_A$-terms depending on
$\mu_f$ cancel there, while the remaining NLL terms are
\beq
 C_A \left[\ln{\frac{s}{\mgl^2-t}} + \ln{\frac{s}{\mgl^2-u}} \right]\,.
\eeq
From the $\bom{K}$-operators in Eq.\ (\ref{eq:2.45}), we get in addition the NLL contributions
\beq
 C_A \left[\ln\frac{\mgl^2}{\mgl^2-t}
 +  \ln{\frac{\mgl^2}{\mgl^2-u}} +2 \right]\,,
\eeq
which together correctly reproduce the contribution from the soft anomalous dimension in
Eq.~\eqref{eq:2.38} and Eq.\eqref{eq:2.48}.

Having computed the resummed and the perturbatively expanded results
in Mellin space, we must multiply them with the $N$-moments of the PDFs according
to Eq.\ (\ref{eq:HadFacN}) and perform an inverse Mellin transform
\bea
 \label{eq:2.51}
 M^2{\d\sigma_{AB}\over\d M^2}(\tau)&=&{1\over2\pi i}\int_{{\cal C}_N}\d N 
 \tau^{-N} M^2{\d\sigma_{AB}(N)\over \d M^2}
\eea
in order to obtain the hadronic cross section as a
function of $\tau=M^2/S$. Special attention must be paid to the
singularities in the resummed exponents $G_{ab}^{(1,2)}$, which are situated at
$\lambda=1/2$ and are related to the Landau pole of the perturbative coupling
$a_s$. To avoid this pole as well as those in the Mellin moments of the PDFs
related to the small-$x$ (Regge) singularity $f_{a/A}(x,\mu_0^2)\propto x^\alpha
(1-x)^\beta$ with $\alpha<0$, we choose an integration contour ${\cal C}_N$
according to the {\em principal value} procedure proposed in
Ref.~\cite{Contopanagos:1993yq} and the {\em minimal prescription} proposed in
Ref.~\cite{Catani:1996yz}. We define two branches
\bea
  {\cal C}_N:~~ N=C+ze^{\pm i\phi}~~{\rm with}~~ z\in[0,\infty[,
  \label{eq:IT:Nbra}
\eea
where the constant $C$ is chosen such that the singularities of the $N$-moments
of the PDFs lie to the left and the Landau pole to the right of the
integration contour. Formally the angle $\phi$ can be chosen in the range
$[\pi/2,\pi[$, but the integral converges faster if $\phi>\pi/2$.

The Mellin moments of the PDFs are obtained by fitting to the parameterisations
tabulated in $x$-space the functional form used by the MSTW collaboration \cite{Martin:2009iq}
\beq
 f(x) = A_0\, x^{A_1}\,  (1 - x)^{A_2} \, \left(1 + A_3 \, \sqrt{x} + A_4 \, x + A_5 \, x^{\frac{3}{2}}\right) + A_6\, x^2 + A_7\, x^{\frac{5}{2}} \,,
\eeq
which has the advantage that it can be transformed analytically with the result
\bea
F(x) &=& A_{0} \, \Gamma\left(y\right) \, \mathrm{B'}\left(A_1 + N, y\right) + A_3 \, \mathrm{B'}\left(A_1 + N + \frac{1}{2}, y\right) + A_4 \mathrm{B'}\left(A_1 + N + 1, y\right)\nonumber \\ 
&+& A_5 \mathrm{B'}\left(A_1 + N + \frac{3}{2}, y\right) + A_6\mathrm{B'}\left(A_1 + N + 2, y\right) + A_7 \mathrm{B'}\left(A_1 + N + \frac{5}{2}, y\right) \,.
\eea
Here, $y = A_2 + 1$ and $\mathrm{B'}(x,y) = \mathrm{B}(x,y)/\Gamma(y) = \Gamma(x)/\Gamma(x + y)$.
We have verified that we obtain good fits not only for the MMHT2014NLO118 \cite{Martin:2009iq},
but also for the CT14NLO fits \cite{Dulat:2015mca} up to large values of $x$ 
and for all typical factorisation scales, 
even though the latter are obtained with an ansatz including an exponential function.
The fit to the NNPDF$\_$30$\_$nlo$\_$as$\_$0118 PDFs \cite{Ball:2014uwa} is slightly less
stable in the large-$x$ region. They will therefore only be used for estimates of the PDF
uncertainty. We compute in this case first an (approximately PDF-independent)
$K$-factor, {\it i.e.}~the ratio of NLL+NLO over NLO cross sections, using stable ({\it e.g.}~CT14NLO)
PDFs and then multiply with it the NLO calculation convoluted with NNPDFs directly in $x$
space.

\section{Numerical results}
\label{sec:3}

We now turn to our numerical results. For our calculations, we have used the
Particle Data Group (PDG) values for the Standard Model parameters, in particular
for the value of the strong coupling constant at the $Z$-pole $\alpha_s(M_Z)$
\cite{Agashe:2014kda}. In our LO and NLO/NLL+NLO calculations, it is evaluated in
the one- and two-loop approximations, respectively, with five active quark flavours.
All light quarks including the bottom quark are taken as massless. The top quark is
decoupled. Its (pole) mass enters only in the gluino self-energy and has little
numerical influence on the production cross sections.

\subsection{Benchmark scenario}
\label{sec:3.1}

Our results are given for a specific phenomenological MSSM (pMSSM)
benchmark scenario with 13 free parameters.
These parameters are listed together with the corresponding fitted numerical values in
Tab.~\ref{tab:1}.
\begin{table}
 \centering
 \begin{tabular}{|ccc|cc|cccccccc|}
 \hline
 $\tan\beta$ & $\mu$ & $m_A$ & $M_1 $ & $M_3$ & $M_{Q_{1,2}}$ & $M_{Q_3}$ & $M_{U_{1,2}}$ & $M_{U_3}$ & $M_{D_{12}}$ & $M_{D_3}$ & $M_{L}$ & $A_f$ \\
 \hline
 21 & \!\!\!773\!\!\! & 1300 & \!\!315 & \!\!1892 & \!\!2288 & \!\!425 & \!\!1758 & \!\!2754 & \!\!552 & \!\!714 & \!\!1553 & \!\!-2200 \\
 \hline
 \end{tabular}
 \caption{Higgs sector and soft SUSY breaking parameters in our pMSSM-13 benchmark model.
 All values except the one for $\tan\beta$ are in GeV.}
 \label{tab:1}
\end{table}
Our scenario is inspired by the benchmark point II of Ref.\ \cite{DeCausmaecker:2015yca}, that
has been obtained with a Markov Chain Monte Carlo (MCMC) scan using also PDG values for
the Standard Model parameters. This 19-parameter scan has been performed with a focus on
non-minimal flavour violation (NMFV). It therefore included seven flavour-violation parameters
and checked in particular the most stringent flavour-changing neutral current (FCNC) constraints
from rare $B$- and $K$-decays. Since we are not interested in NMFV, we set these parameters
all to zero. This reduces the SUSY contributions to the rare meson decays. To compensate
for the reduced mass splitting in the top squark sector and obtain a Higgs-boson mass compatible
with the measured value, we have instead changed the sign and increased the absolute value of the
trilinear coupling $A_f$. We then still obtain a neutralino lightest SUSY particle
and in addition a light top
squark, which leads to a viable dark matter candidate and allows
in general for sufficient stop coannihilation to reproduce the observed dark matter relic density
\cite{Klasen:2015uma,Harz:2012fz,Harz:2014tma,Harz:2014gaa,Harz:2016dql}. While we continue
to impose the GUT relation between the bino and wino mass parameters, $M_1\simeq M_2/2$,
we allow the gluino mass parameter $M_3$ to vary independently, which brings us to $19-7+1=13$
free parameters. For our default scenario, we still impose the GUT relation for $M_3\simeq6M_1$.

\begin{figure}
 \centering
 \includegraphics[width = 0.9\textwidth]{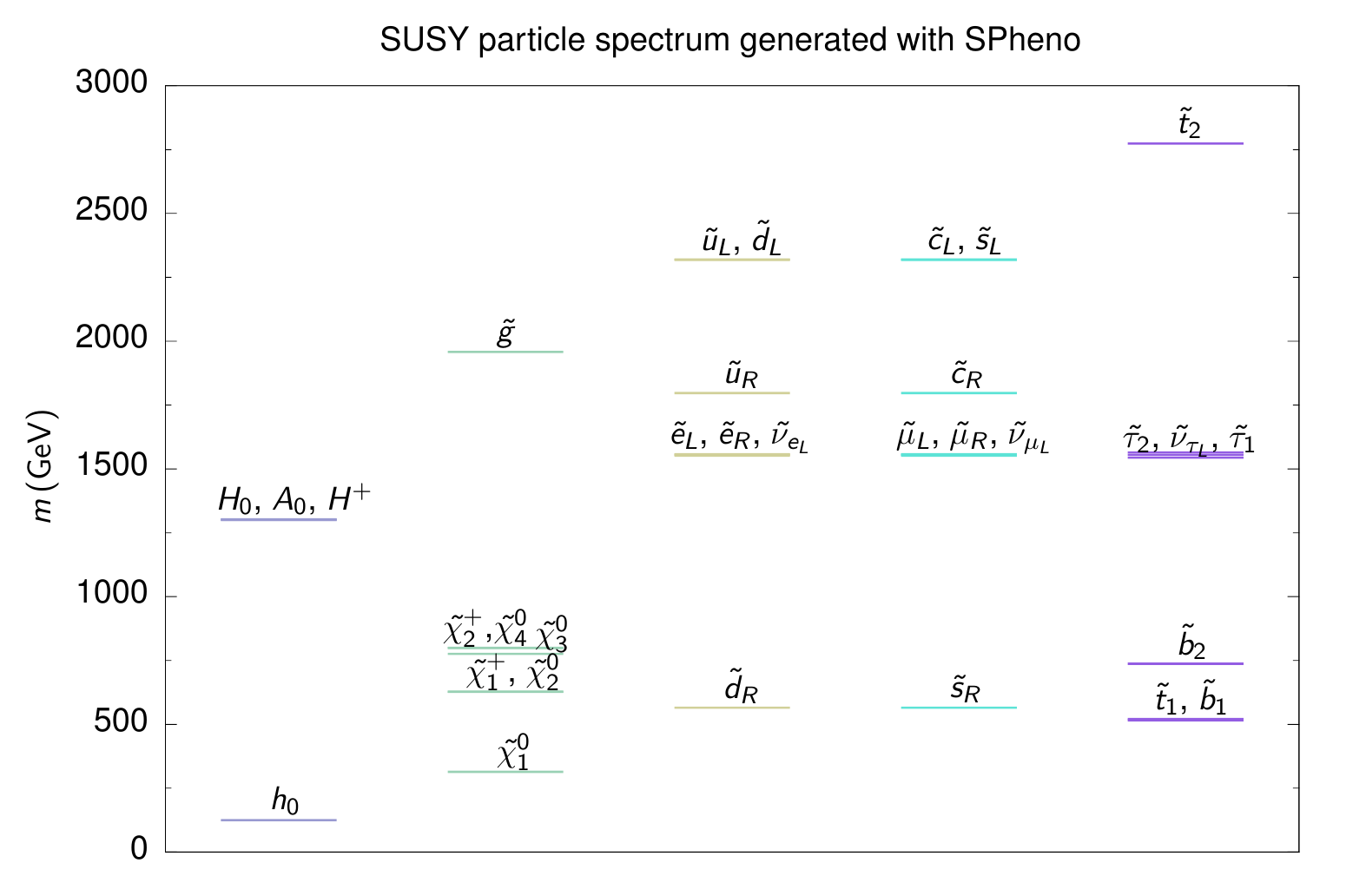}
 \caption{Visualised mass spectrum for the benchmark point defined in Tab. \ref{tab:1}.
 Particles are grouped in Higgs particles, gauginos, first-, second- and third-generation
 sfermions (left to right).}
 \label{fig:03}  
\end{figure}
The physical SUSY mass spectrum is obtained with {\sc SPheno 3.37}~\cite{Porod:2011nf} and shown
schematically in Fig.\ \ref{fig:03}. Apart from the FCNC, the Higgs-boson and neutralino
dark matter data, also the observed value of the anomalous magnetic moment of the muon,
which is unaffected by the gluino mass, is reproduced in this scenario.
In addition, it satisfies
the increasingly stringent constraints that are imposed on the masses of the
SUSY particles from direct search results at the LHC.
For example, with the 2015 data from Run~II, ATLAS and CMS exclude
gluino masses up to 1400 and 1280~GeV, assuming masses of the lightest neutralino of up to 600
and 800 GeV, respectively~\cite{Aad:2016jxj,CMS:2015vqc}. Mass-degenerate light charginos and
second-lightest neutralinos produced electroweakly have been excluded at Run~I up to 465 and
720~GeV in the case of massless lightest
neutralinos~\cite{Aad:2014vma,Khachatryan:2014qwa}.
These exclusion limits are however not valid within the general pMSSM, as they
have been obtained
assuming direct production cross sections and simplified decay scenarios.

As we have already mentioned in the previous section, our calculations allow for arbitrary
squark mixings and mass eigenstates in the appearing couplings and propagators. The mixing of
squark interaction eigenstates is numerically relevant only for third-generation (and in
particular top) squarks. Since the top and bottom quark PDFs are small, the mixing influences
predominantly the gluino self-energy diagram with little (below the percent level) numerical
effect. In contrast, the masses of first- and second generation squarks in our scenario span
a large range from about 600 to 2300 GeV. Averaging over these masses leads to cross sections
that are almost a factor of two larger. This is already true at LO, since the dominant effect
comes from squark propagators already present there. If the LO calculations are performed
without averaging and corrected by a mass-averaged $K$-factor, as it is, e.g., done in
{\sc Prospino} \cite{Beenakker:1996ed}, differences of about 5\% remain. These differences
originate in particular from intermediate squarks at NLO, which in the general case can
sometimes be on-shell, while they cannot be on-shell in the mass-averaged case.

\subsection{Invariant mass distribution}
\label{sec:3.2}

The associated production of a gluino and the lightest neutralino will be difficult to
observe at the LHC, as the latter escapes directly undetected. It is therefore more promising
to study the associated production of a gluino with the second-lightest neutralino (or the
lightest chargino of often equal mass), since it will decay into an additional $Z$ (or $W$)
boson, whose leptonic decay products will then lead to an identifiable signal and better
background suppression. As our default PDFs, we use CT14NLO at NLO and NLL+NLO and CT14LL
together with the one-loop approximation for $\alpha_s$ at LO (see above)~\cite{Dulat:2015mca}.

\begin{figure}
 \centering
 \includegraphics[width = 0.9\textwidth]{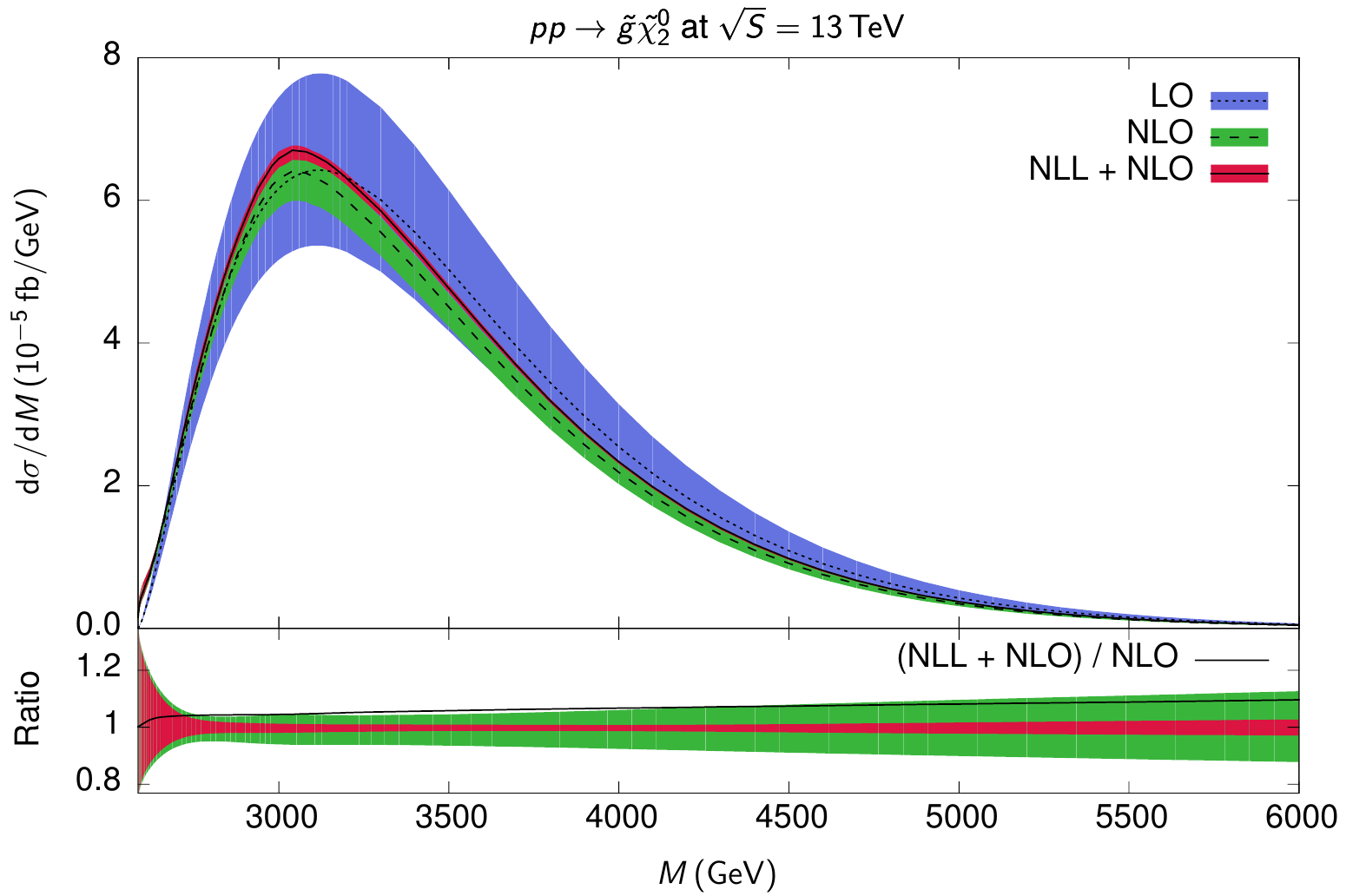}
 \caption{Upper panel: Invariant mass distribution for the process $pp\rightarrow\tilde{g}
 \tilde{\chi}_2^0$ at the LHC with a centre-of-mass energy of $\sqrt{S} = 13$ TeV 
 at the LO (blue),
 NLO (green) and NLL+NLO (red) accuracy. Lower panel: Corresponding relative scale uncertainties and the
 NLL+NLO/NLO $K$-factor (black line).}
 \label{fig:04}
\end{figure}

In the upper panel of Fig.~\ref{fig:04}, we show the invariant-mass distribution
given by Eq.~\eqref{eq:2.51}, for
the production of a gluino with a mass of 1892~GeV
and a second-lightest neutralino with a mass of 630~GeV,
where both masses have been chosen such that they lie beyond current LHC limits even in simplified
scenarios. The cross sections peak at about 3.2 TeV and then fall off towards higher invariant
masses $M$. Due to additional radiation, the maximum is shifted from LO (blue) to NLO (green) and
NLL+NLO (red) towards slightly smaller values of $M$. At the same time, the scale uncertainties
(shaded bands) are significantly reduced from LO to NLO and then again to NLL+NLO. The second
reduction is more clearly visible in the lower panel of Fig.\ \ref{fig:04}, where it amounts to
a change from $\pm12\%$ to $\pm3\%$ at high invariant masses. The scale errors have been
obtained here in the usual way from individual variations of the factorisation and
renormalisation scales by a factor of two about the average mass of the two produced final-state
particles, $(\mgl+m_{\tilde{\chi}_2^0})/2$, excluding relative factors of four. In the high-mass
region, the corrections from threshold resummation at NLL increase the central NLO cross section
by up to 10\% (black line) and more closer to the threshold.

\subsection{Scale uncertainty of the total cross section}
\label{sec:3.3}

\begin{figure}
 \centering
 \includegraphics[width = 0.9\textwidth]{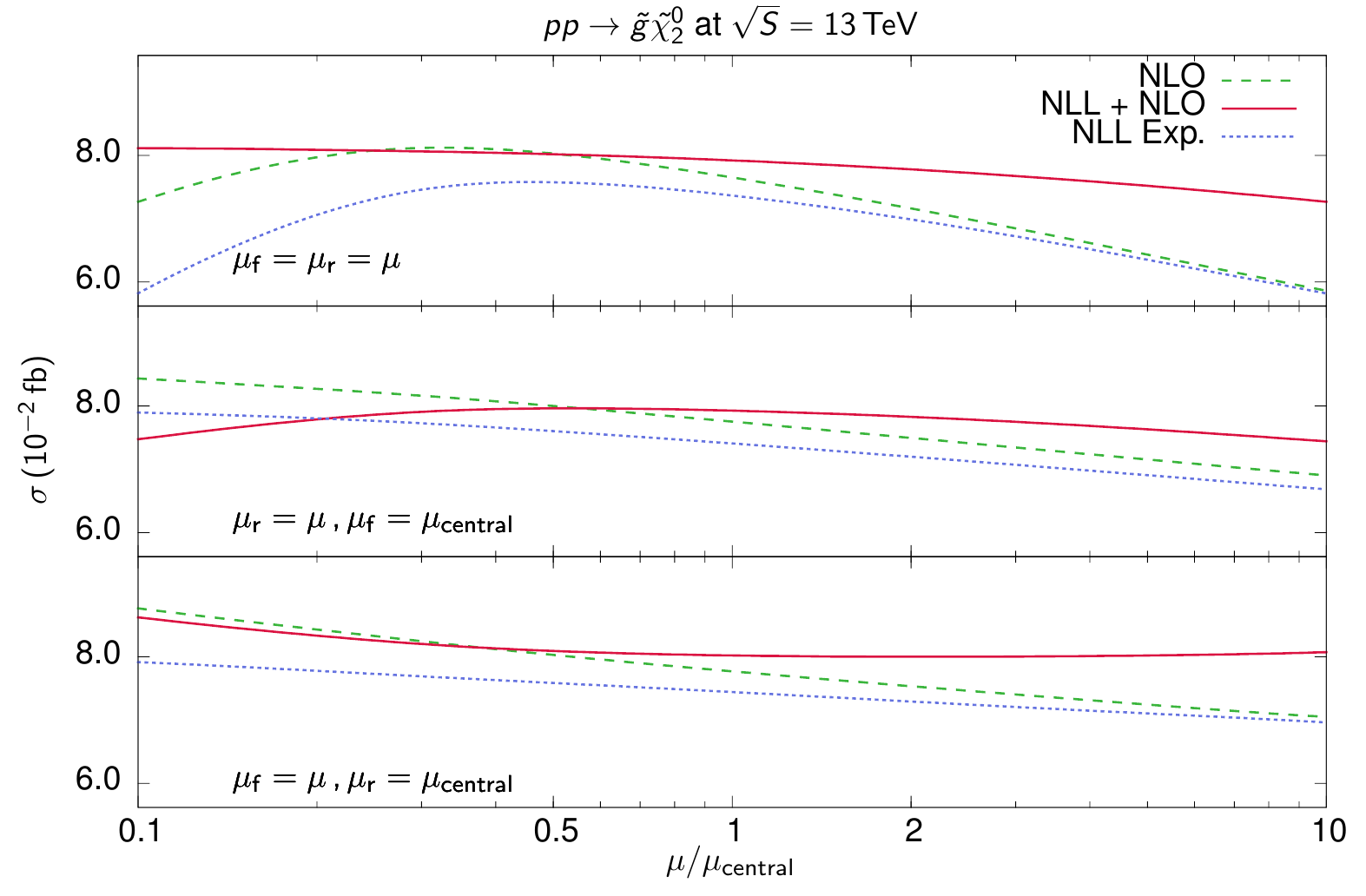}
 \caption{Total cross section for the process $pp\rightarrow\tilde{g}\tilde{\chi}_2^0 $ at the
 LHC with a centre-of-mass energy of $\sqrt{S} = 13$ TeV at NLO (green dashed curve), NLL+NLO
 (red full curve) and after the re-expansion of the NLL result to NLO (blue dotted curve). We vary
 the renormalisation scale and factorisation scale together (upper panel), only the
 renormalisation scale with fixed factorisation scale (central panel) and {\it vice versa} (lower
 panel).}
 \label{fig:05}
\end{figure}

After integrating over the invariant mass $M$ in Eq.~\eqref{eq:2.51}, we obtain the total
production cross section for a gluino and a second-lightest neutralino. For a process that
depends already at LO on the strong coupling constant, one expects the significant
(approximately logarithmic) scale dependence to be already reduced at NLO. This is clearly
visible in Fig.\ \ref{fig:05}, where the NLO result (green dashed curve) shows the characteristic
maximum at approximately half the central renormalisation and factorisation scale (upper panel).
The scale dependence is further reduced at NLL+NLO (red full curve), as it was already the case
for the invariant-mass distribution (see above). When the NLL+NLO result is re-expanded to NLO
(blue dotted curve), it becomes a good approximation to the full NLO result in particular
for large scale choices, when the logarithmic terms dominate the cross section. This is also
true when the renormalisation (central panel) and factorisation scales (lower panel) are
varied individually and not together. The lower two panels also demonstrate nicely the interplay
of the renormalisation and factorisation scale behaviour in the NLO and NLL+NLO cross sections,
that together produce the stabilised behaviour in the upper panel with a large plateau
in particular at NLL+NLO. We have verified that we obtain similar results also for larger gluino
masses of, {\it e.g.}, 3~TeV.

\subsection{Gluino mass dependence of the total cross section}
\label{sec:3.4}

\begin{figure}
 \centering
 \includegraphics[width = 0.9\textwidth]{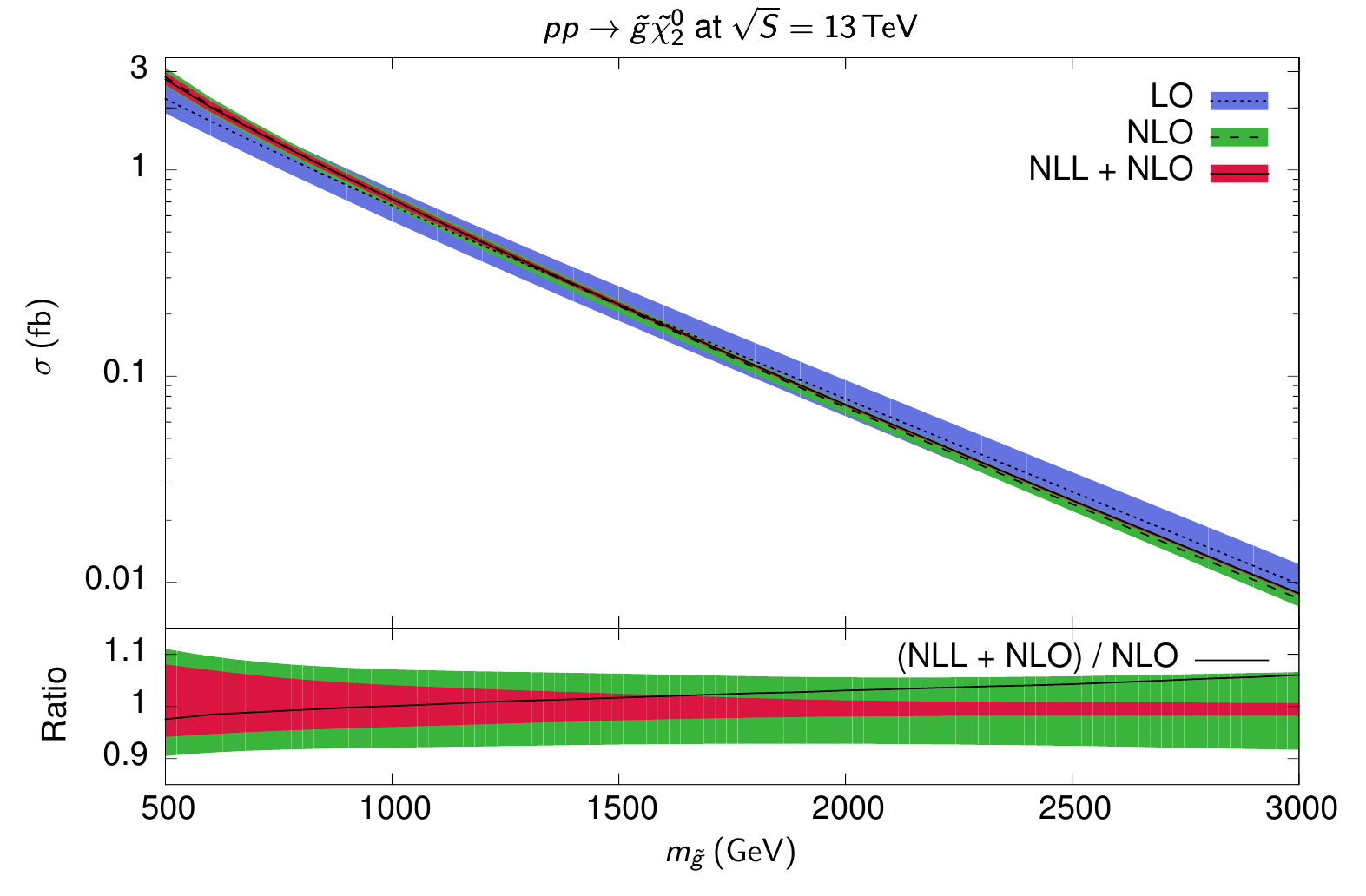}
 \caption{Upper panel: Total cross section for the process $pp\rightarrow\tilde{g}
 \tilde{\chi}_2^0 $ at the LHC with a centre-of-mass energy of $\sqrt{S} = 13$ TeV in LO
 (blue), NLO (green) and NLL+NLO (red) as a function of the gluino mass.
 Lower panel: Corresponding relative scale uncertainties and the NLL+NLO/NLO $K$-factor
 (black line).}
 \label{fig:06}
\end{figure}

Since the gluino mass is unknown, it is interesting to compute the total cross section
for associated gluino-neutralino production as a function of the gluino or gaugino
mass. The gluino mass dependence is shown in the upper panel of Fig.\ \ref{fig:06}
and in tabular form in Tab.\ \ref{tab:c1} in App.\ C.
As expected, the cross section falls steeply with the gluino mass from 3 to 0.01 fb
in the range $\mgl\in[500, 3000]$~GeV. With LHC luminosities of currently a few fb$^{-1}$
and in the near future a few 100 fb$^{-1}$ at $\sqrt{S}=13$ TeV, these cross sections will
soon be observable. In the high-luminosity phase of the LHC, expected to collect up to
3000 fb$^{-1}$, even larger gluino masses can be reached that may kinematically no longer
be accessible in the strong production of gluino pairs. 

As the lower panel of Fig.\ \ref{fig:06} shows, the NLO scale uncertainty on the total
cross section of $\pm10$\% at 500 GeV decreases only slightly towards higher gluino masses.
This is in sharp contrast to the NLL+NLO prediction, that has already a smaller scale
error of $\pm7$\% at 500 GeV and that becomes much more reliable with an error of only a few
percent at large gluino masses. At the same time, the NLO cross section is increased at NLL+NLO
by 7 (black line) to 20\% for gluino masses of 3 to 6 TeV, respectively.

\subsection{Gaugino mass dependence of the total cross section}
\label{sec:3.5}

\begin{figure}
 \centering
 \includegraphics[width = 0.9\textwidth]{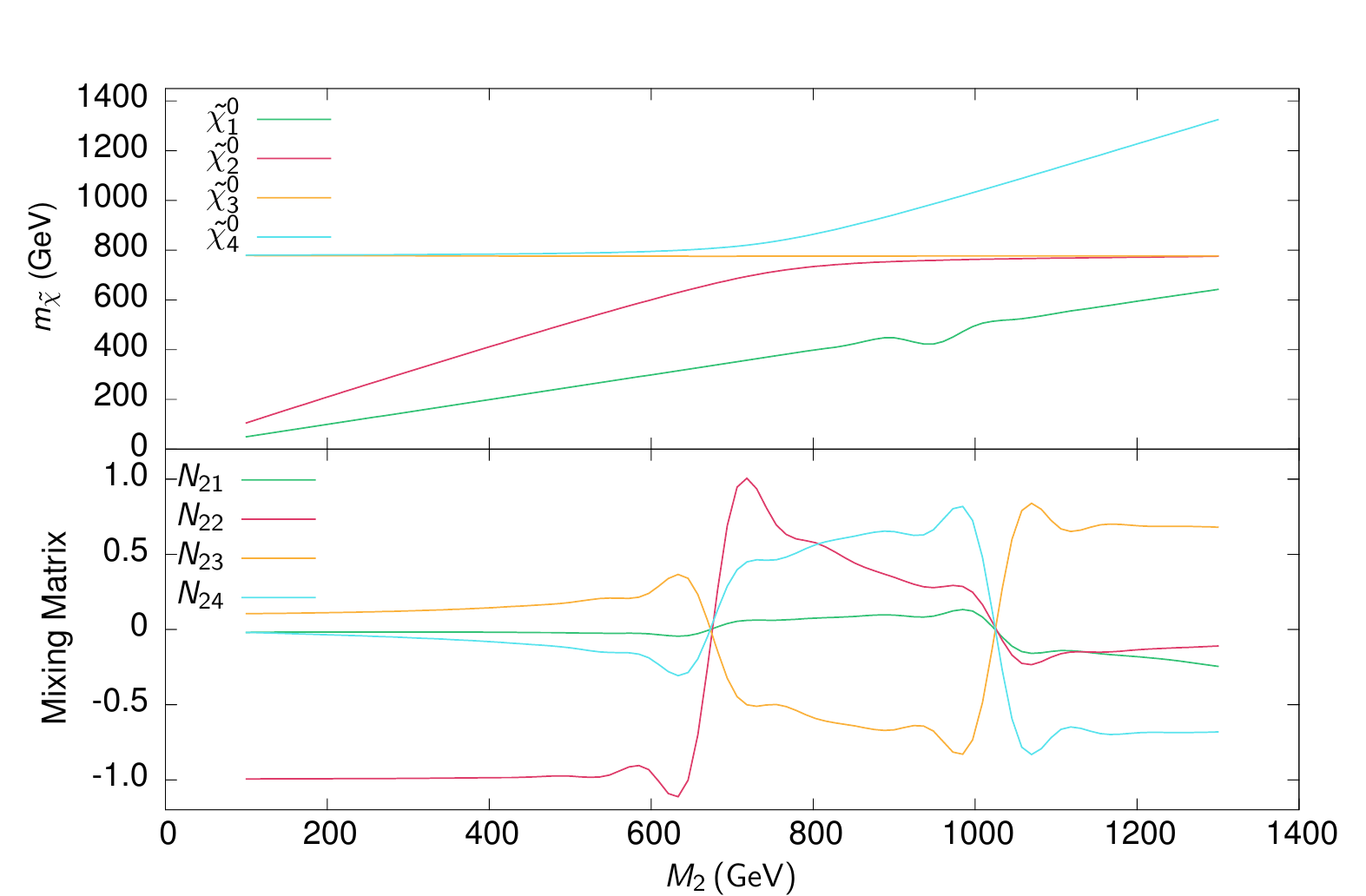}
 \caption{Dependence of the neutralino mass eigenvalues (upper panel) and mixing matrix
 elements of the second-lightest neutralino (lower panel) on the wino and bino mass parameters
 $M_2 \simeq 2 \, M_1$. All other Higgs and soft SUSY-breaking parameters have been kept fixed and
 set to the values of our benchmark scenario.}
 \label{fig:07}
\end{figure}
Similarly to the gluino mass dependence, the total cross section for gluino-gaugino
associated production decreases with the gaugino mass. Since the mass of the
second-lightest neutralino (and the almost identical one of the lightest chargino)
is a dependent physical mass obtained after diagonalisation of the neutralino (or
chargino) mass matrix with the mixing matrix $N$, we vary instead the bino mass
parameter $M_1$, which fixes
immediately also the wino mass parameter $M_2$ through the GUT relation $M_2 \simeq 2 \, M_1$.
As one can see in the upper panel of Fig.\ \ref{fig:07}, the mass eigenvalue of the
second-lightest neutralino $\tilde{\chi}_2^0$ increases linearly with $M_2$ up to $M_2=\mu
=773$ GeV, where a typical avoided crossing occurs. At higher values of $M_2$, it is
the mass eigenvalue of $\tilde{\chi}_4^0$ that depends linearly on $M_2$, while the
mass of $\tilde{\chi}_2^0$ stays constant. Accordingly, the decomposition of $\tilde{\chi}_2^0$
changes from wino-type (large mixing matrix element $N_{22}$, red curve) to higgsino-type
(large mixing matrix elements $N_{23}$ and $N_{24}$, yellow and blue curves).
Both features will of course influence the production cross section of the process
$pp\to\tilde{g}\tilde{\chi}_2^0$. 

\begin{figure}
 \centering
 \includegraphics[width = 0.9\textwidth]{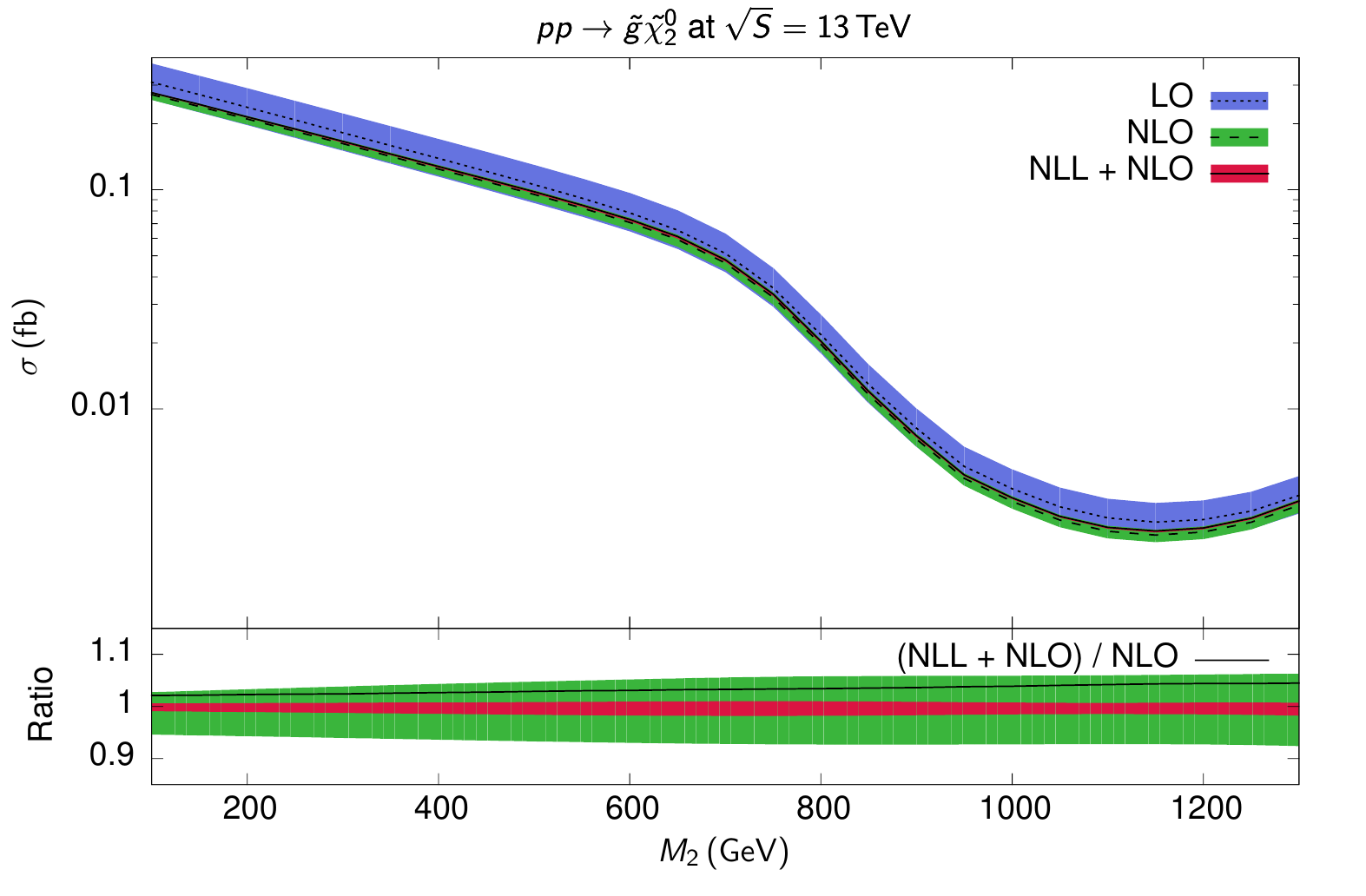}
 \caption{Upper panel: Total cross section for the process $pp\rightarrow\tilde{g}
 \tilde{\chi}_2^0 $ at the LHC with a centre-of-mass energy of $\sqrt{S} = 13$ TeV in LO
 (blue), NLO (green) and NLL+NLO (red) as a function of the wino mass parameter $M_2$. 
 The bino mass parameter has been varied simultaneously using the GUT relation $M_1\simeq M_2/2$.
 Lower panel: Corresponding relative scale uncertainties and the NLL+NLO/NLO $K$-factor
 (black line).}
 \label{fig:08}
\end{figure}
The dependence of the cross section on the wino mass parameter is shown in Fig.\
\ref{fig:08} and in tabular form in Tab.\ \ref{tab:c2} in App.\ C.
As expected, it falls with $M_2$ as long as the physical mass of
$\tilde{\chi}_2^0$ changes. For $M_2>\mu$, the neutralino becomes higgsino-like and couples
mostly via quark Yukawa couplings. Since the heavy quark PDFs in the proton are small,
the cross section starts to fall even faster than before despite the fact that
the gaugino mass remains constant and the available phase space no longer changes.
Interestingly, the cross section increases again somewhat for $M_2>1150$~GeV,
which can be explained with a slightly increasing bino component (see Fig.\ \ref{fig:07}).
The scale uncertainty (shaded bands) is drastically reduced from LO (blue) to NLO (green),
in particular at low values of $M_2$. However, the NLO scale dependence increases with
$M_2$ up to $M_2=\mu=773$~GeV due to the rising contribution of the large logarithms.
This can be seen more clearly in the lower panel of Fig.~\ref{fig:08}. 
Beyond this mass, the scale dependence remains constant as expected. A similar
trend is observed in the NLL+NLO scale dependence (red), albeit at an again much
lower level. To be specific, it rises only from $\pm1$ to $\pm3$\% compared to a
scale dependence at NLO that rises from $\pm3$ to $\pm6$\%. The $K$-factor (black
line) increases also with $M_2$ from 1.02 to 1.07.

\subsection{Parton density uncertainty of the total cross section}
\label{sec:3.6}

While the scale uncertainty is expected to be reduced due to the resummation of
large logarithms, the PDF uncertainty is normally not improved by this procedure.
It is usually estimated by propagating the experimental uncertainties on the fitted
data points through to the PDFs ({\it e.g.}~linearly via a Hessian method) and leads
to the production of orthogonal eigenvector PDF sets corresponding to a 90\% confidence
level. This method is employed by the CTEQ and MMHT collaborations and has been found
to produce comparable results to the more intricate Lagrange multiplier method
\cite{Dulat:2015mca,Martin:2009iq}. The uncertainty on the cross section is then obtained
from 
\bea
 \Delta\sigma_{\rm PDF+}&=&\sqrt{\sum_{i=1}^{n}\left[\max\left(\sigma_{+i}-\sigma_0,
 \sigma_{-i}-\sigma_0,0\right)\right]^2},\\
 \Delta\sigma_{\rm PDF-}&=&\sqrt{\sum_{i=1}^{n}\left[\max\left(\sigma_0-\sigma_{+i},
 \sigma_0-\sigma_{-i},0\right)\right]^2},
\eea
where $n=28$ and 25 is the number of eigenvector directions in the CT14 and MMHT2014
fits, respectively. Since the MMHT PDF sets are only available for 68\% and not
90\% confidence level, the corresponding error must be multiplied by the standard
factor of 1.645 for compatibility. The NNPDF collaboration uses instead a Monte Carlo method,
where the PDF uncertainty is obtained by sampling the available replicas
\cite{Ball:2014uwa}. Since PDF uncertainties are usually
not produced for LO fits and since the results are very similar at NLO and NLL+NLO, we
will only study them at the level of NLL+NLO cross sections. For NNPDF, we
estimate the PDF uncertainty using the $K$-factor method described in Sec.\ \ref{sec:2.5}.

\begin{figure}
 \centering
 \includegraphics[width = 0.9\textwidth]{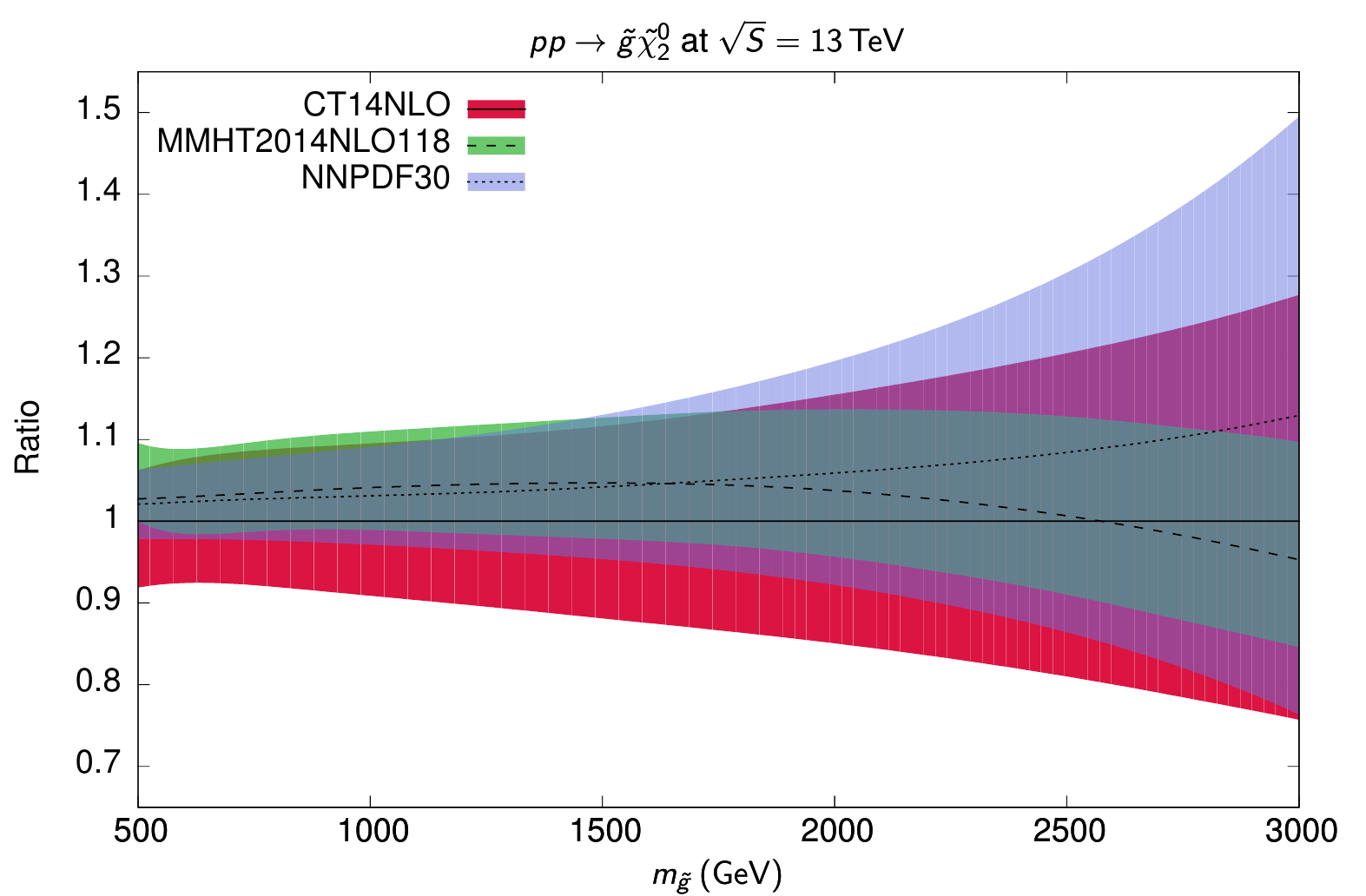}
 \caption{Relative PDF uncertainties of the total cross section for the process $pp\rightarrow
 \tilde{g}\tilde{\chi}_2^0 $ at the LHC with a centre-of-mass energy of $\sqrt{S} = 13$ TeV in
 NLL+NLO. The uncertainties are shown for three different PDFs, CT14NLO (red), MMHT2014NLO118
 (green) and NNPDF30 (blue), as a function of the gluino mass $\mgl$.}
 \label{fig:09}
\end{figure}

The results are shown in Fig.\ \ref{fig:09}. As one can see, the estimates by the three groups
overlap to a large extent. While the three individual PDF uncertainties are still relatively
small and comparable to the scale uncertainty with about $\pm5$\% at small values of the gluino
mass of about 500 GeV (cf.\ Fig.\ \ref{fig:06}), they increase rather than decrease with $\mgl$
and reach a level of $\pm20$\% at 3 TeV. This is of course due the fact that the PDFs are much
less constrained at large than at intermediate values of the parton momentum fraction $x$.
Compared to the central prediction with CT14, those with MMHT2014 and NNPDF30 lie systematically
higher by a few percent. The increase of the uncertainty towards larger $x$ is less pronounced
in MMHT and more pronounced in NNPDF. It will therefore be interesting to study the impact of
threshold-improved PDFs in future work, as it was done for squark and gluino production
\cite{Bonvini:2015ira,Beenakker:2015rna}. It is important to note that the scale uncertaintites
computed in the previous sections and the PDF uncertainty computed in this section are
independent and therefore usually added in quadrature for a reliable estimate of the
total theoretical uncertainty.

\section{Conclusion}
\label{sec:4}

We have presented in this paper a threshold resummation calculation at the NLL+NLO accuracy
for the associated production of gluinos and gauginos at the LHC. This process
is of intermediate strength compared to the strong production of gluino (and
squark) pairs and the electroweak production of gaugino (and slepton) pairs.
It can in particular become relevant should the gluinos prove to be too heavy
to be pair-produced at the LHC. This situation would not be unexpected, if
one takes the GUT relation between the gaugino masses seriously, which predicts
$M_1=M_2/2=M_3/6$ after renormalisation group running at the weak scale.
Lighter gluinos, {\it e.g.}~with a possible cosmological impact through their
coannihilation with gauginos, typically require non-minimal assumptions such
as non-universal gaugino masses or vector-like supermultiplets \cite{Profumo:2004wk,%
Nagata:2015hha,Ellis:2015vna,Nath:2016kfp}. Conversely, associated gluino-gaugino
production could also become phenomenologically important in the less likely case
that the gauginos lie beyond the kinematic reach of the LHC, so that {\it e.g.}~the
gravitino becomes the lightest SUSY particle and its associated production with
(relatively light) gluinos an interesting search channel \cite{Klasen:2006kb}.

Our investigations required the (re-)calculation of the full NLO corrections,
which we generalised to the case of non-degenerate squark masses, and of the
process-dependent soft anomalous dimension and hard matching coefficients, which
we could show to be consistent with each other. For a typical benchmark scenario,
obtained in a recent MCMC fit of the Higgs boson, FCNC, muon magnetic
moment and LHC data, we presented numerical predictions for the invariant-mass
distribution and the total cross section as a function of the gluino or
gaugino mass. The resummation of the NLL contributions increased the NLO cross
sections at large invariant mass by up to 10\% and stabilised them dramatically
with respect to the scale dependence. As expected, the PDF uncertainty was, however,
not reduced. It will therefore be interesting to study the impact of threshold-improved
PDFs in the future.

Numerical predictions for other SUSY scenarios are available from the authors upon request.
The calculation will also be included in the next release of the public code
{\sc Resummino}~\cite{Fuks:2013vua}.
Its application to the associated production of gluinos and gravitinos
would not only require the inclusion of the additional $s$-channel gluon exchange, but also
a full NLL+NLO calculation for the gluon-initiated diagrams. The NLL+NLO calculation for the
associated production of squarks and gauginos is in progress and will be presented elsewhere.

\section*{Acknowledgments}
M.R.\ thanks V.\ Theeuwes for useful discussions. This work has been supported by
the BMBF under contract 05H15PMCCA, by the CNRS under contract PICS 150423 and the
Th\'eorie-LHC-France initiative of the CNRS (IN2P3/INP).

\appendix
\section{Coupling conventions}
\label{sec:a}

\begin{figure}[h]
 \centering
 \includegraphics[width = \textwidth]{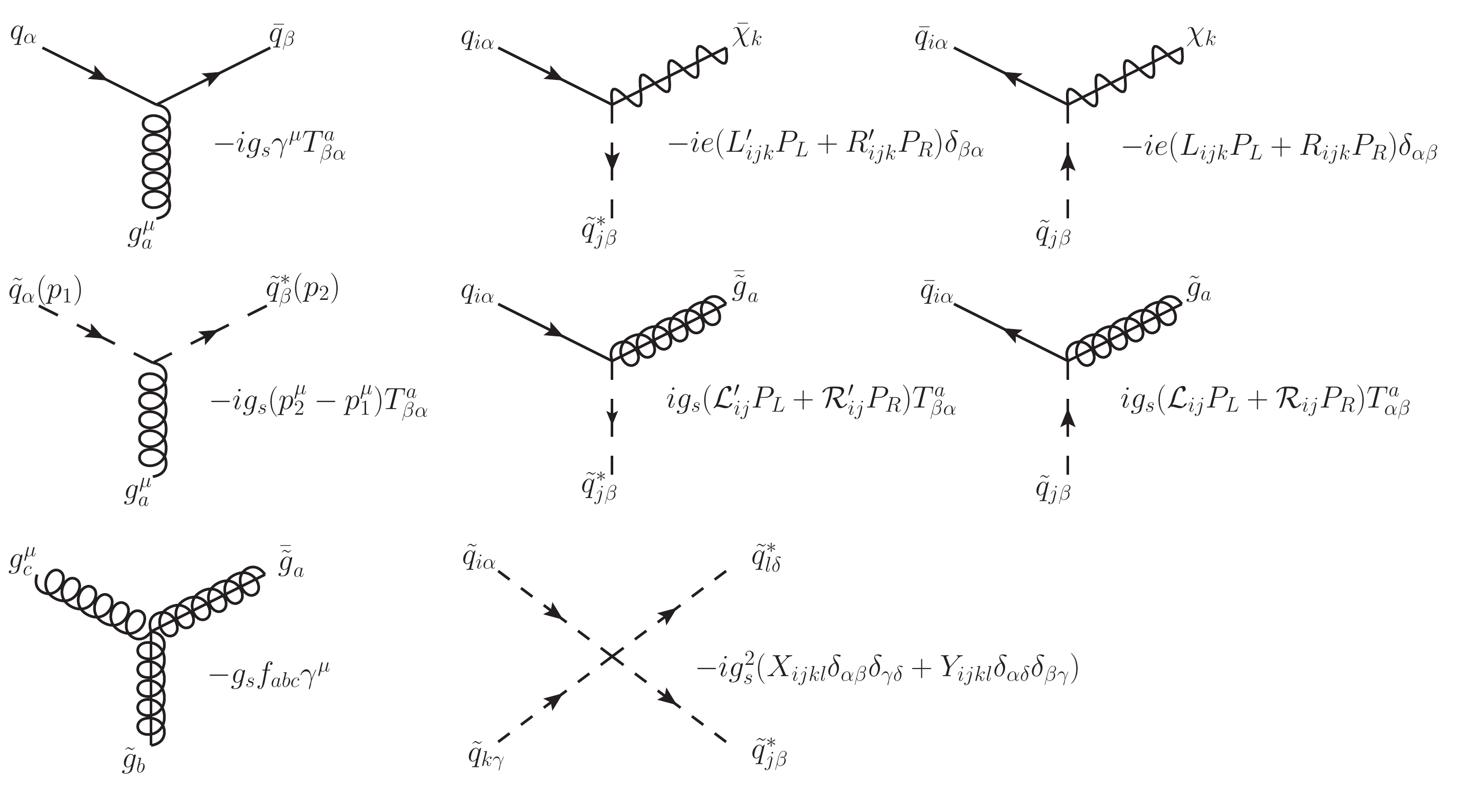}
 \caption{Interaction vertices appearing in the associated production of gluinos and gauginos
 at LO and NLO. All momenta are ingoing, and arrows describe charge/fermion flow.}
 \label{fig:FeynmanRules}
\end{figure}

The conventions for the couplings appearing in our calculations are defined in
Fig.~\ref{fig:FeynmanRules}. The electromagnetic and (renormalisation scale dependent) strong
coupling constants are denoted by $e$ and $g_s(\mu_r)$,
respectively. $T^a_{\beta\alpha}$ and $f_{abc}$ are SU(3) colour matrices in
the fundamental representation and structure constants,
$\gamma^\mu$ and $P_{L,R}$ Dirac matrices and chirality projection operators.
The latter are associated with
generic MSSM coupling constants $L^{(')}$, $R^{(')}$,
${\cal L}^{(')}$ and ${\cal R}^{(')}$ which involve squark and gaugino mixing matrices and
can be found together with the quartic squark couplings $X$ and $Y$ in Refs.\
\cite{Haber:1984rc,Gunion:1984yn}.

%


\section{Soft anomalous dimension}
\label{sec:b}

If the calculation of the (modified) soft anomalous dimension is performed in the axial
gauge with a gauge vector $n^\mu$, it is given by
\beq
 {\bar \Gamma}_{ab \rightarrow ij,IJ} = \Gamma_{ab \rightarrow ij,IJ} - \frac{\alpha_s}{2 \pi}
\sum \limits_{k = {\{a,b\}}} C_k \left( 1-\ln{\left(2 \frac{(v_k \cdot n)^2}{\vert n \vert^2}\right)} - i \pi \right) \delta_{IJ},
\eeq
where one sums over the two incoming particles and
where $|n|^2=-n^2-i\eps$ \cite{Kidonakis:1997gm,%
Beenakker:2009ha}. The dimensionless vector $v_k$ is given by the momentum of the incoming
massless particle $k$ rescaled by $\sqrt{2/s}$. Here, the soft anomalous dimension has been
modified (subtracted) for the soft functions of the two incoming Wilson lines annihilating into
a colour-singlet, {\it i.e}~the Drell-Yan process, effectively isolating the gauge dependence of a
single line and making the soft functions separately gauge invariant.

\begin{figure}
 \centering
 \includegraphics[width = 0.75 \textwidth]{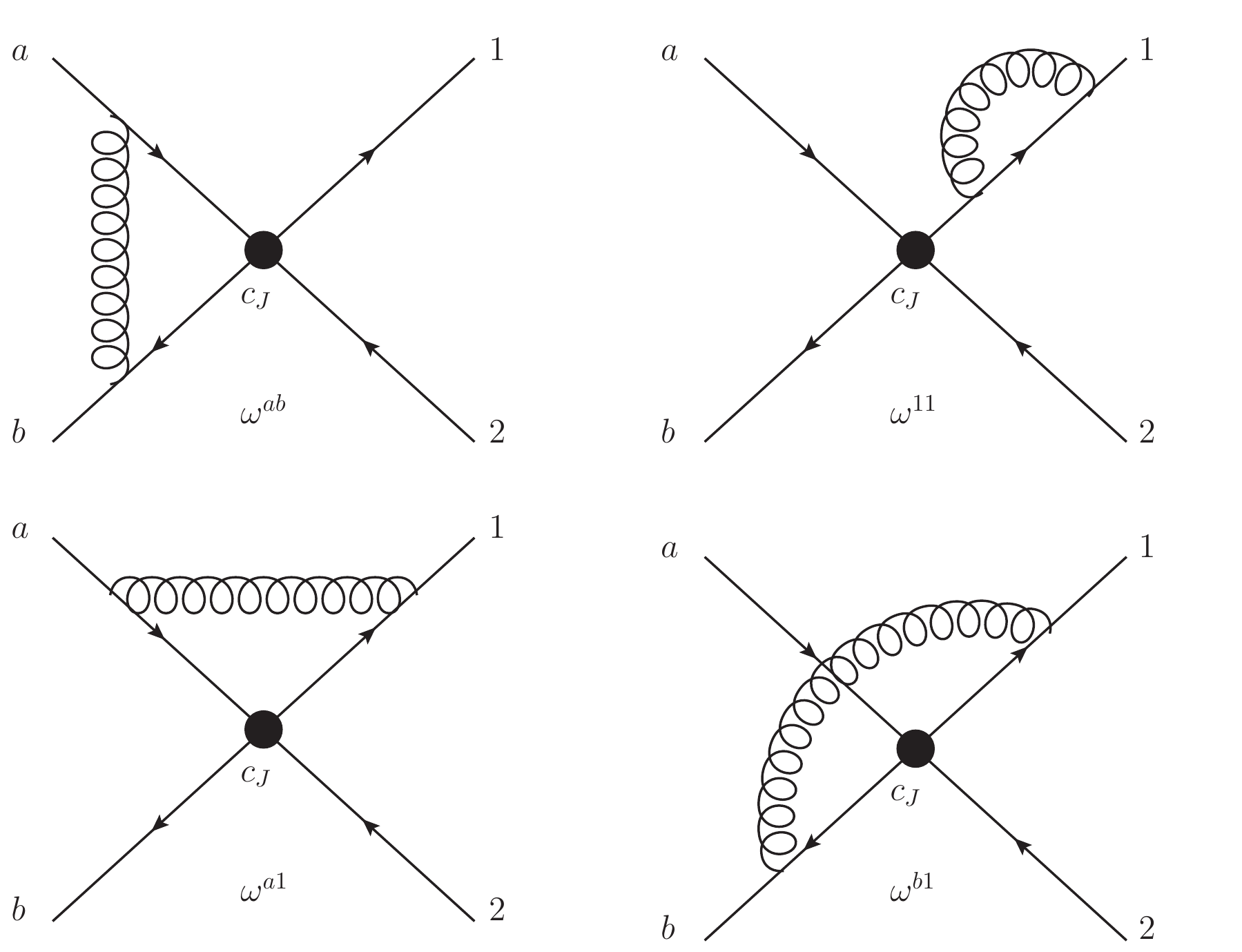}
 \caption{One-loop diagrams contributing to the soft anomalous dimension for the associated
 production of massive colour-octet gluinos and colour-singlet gauginos with momenta $p_1$ and $p_2$ from
 massless colour-triplet quarks and antiquarks with momenta $p_a$ and $p_b$. The self-energy
 contributions of the latter vanish.}
 \label{fig:02} 
\end{figure}

Soft anomalous dimensions are computed from the renormalisation constants $Z_{IJ}$ of Wilson-line
operator products by taking the residues of their ultraviolet poles in $\eps=4-D$ in $D$
dimensions,
\beq
 \Gamma_{ab\to ij,IJ}=-\alpha_s{\partial\over\partial\alpha_s}{\rm Res}_{\eps\to0}
 Z_{ab\to ij,IJ}(\alpha_s,\eps).
\eeq
Here, we only need the one-loop corrections, depicted in Fig.~\ref{fig:02}. If we denote
by $k$ and $l$ the eikonal lines, between which the gluon is spanned, by $C^{kl}_{IJ}$ the colour
mixing factors and by $\omega^{kl}$ the kinematic parts of the one-loop corrections, we obtain
for the correction to the colour basis tensor $c_I$
\beq
 \Gamma_{ab\to ij,IJ}=-\sum_{kl}C^{kl}_{IJ}{\rm Res}_{\eps\to0}\omega^{kl}.
\eeq

We compute these diagrams in an irreducible $s$-channel colour basis with tensors $c_J$, which
one obtains in general after decomposing the reducible initial-state or final-state
multi-particle product representations and which has the advantage of rendering the anomalous
dimension matrices diagonal at threshold~\cite{Kidonakis:1997gm,Beenakker:2009ha}. Since we have
only one coloured final-state particle, the gluino with adjoint colour index $i,i'$, there is
also only one colour basis tensor $c_J={\rm Tr}(T^iT^{i'})=T_F\delta_{ii'}$ with $T_F=1/2$,
similarly to the case of prompt photon production with an associated gluon jet~\cite{Kidonakis:1999hq},
and we can drop the associated indices $I,J$. At LO, it leads to the
colour factor ${\rm Tr}(T^iT^{i})=T_F\delta_{ii}=C_AC_F$ computed in Sec.~\ref{sec:2.1}. At one
loop and after removing the LO colour factors, we obtain for the four diagrams in Fig.~\ref{fig:02}
\bea
 C^{ab}~=~{\Tr (T^iT^jT^iT^j)\over\Tr(T^iT^i)}~=~C_F-{C_A\over2}&\ ,\ &
 C^{a1}~=~{\Tr (T^iT^{i'}T^j)f^{ii'j}(-i)\over\Tr(T^iT^i)}~=~{C_A\over2},\\
 C^{b1}~=~{\Tr (T^iT^jT^{i'})f^{ii'j}(-i)\over\Tr(T^iT^i)}~=~-{C_A\over2}&\ ,\ &
 C^{11}~=~{\Tr (T^iT^{i'})f^{ii''j}f^{i''i'j}(-i)^2\over\Tr(T^iT^i)}~=~C_A,\qquad~~
\nonumber\eea
where $j$ is the colour index of the exchanged gluon.

The kinematic part of the one-loop corrections can be written in a general form as
\bea
 \omega^{kl} &=& g_s^2 \int \frac{\dd^Dq}{(2\pi)^D} \frac{-i}{q^2 + i \epsilon}
 \biggl[ \frac{\Delta_k \Delta_l v_k \cdot v_l}{(\delta_k v_k \cdot q + i \epsilon)
                                                (\delta_l v_l \cdot q + i \epsilon)} \nonumber\\
 &-& \frac{\Delta_k v_k \cdot n}{(\delta_k v_k \cdot q + i \epsilon)} \frac{P}{(n \cdot q)}
  -  \frac{\Delta_l v_l \cdot n}{(\delta_l v_l \cdot q + i \epsilon)} \frac{P}{(n \cdot q)}
  +  n^2 \frac{P}{(n \cdot q)^2} \biggl]\,,
\eea
where $q$ is the loop momentum, $\Delta_{k,l}$ and $\delta_{k,l}$ are signs associated with the
eikonal Feynman rules, and $P$ stands for the principle value \cite{Kidonakis:1997gm,%
Beenakker:2009ha}
\beq
 \frac{P}{(n \cdot q)^\beta} = \frac{1}{2} \left( \frac{1}{(n \cdot q + i \epsilon)^\beta}
 + (-1)^\beta \frac{1}{(-n \cdot q + i \epsilon)^\beta} \right)\,.
\eeq
The integrals can be solved \cite{Kidonakis:1997gm} (also for two coloured final-state particles
with unequal masses \cite{Beenakker:2009ha}) with the results \cite{Vincent}
\bea
 \omega^{ab} &=& S_{ab} \frac{\alpha_s}{\pi \epsilon} \left[ - \ln{\left( \frac{v_a \cdot v_b}{2}\right)} + \frac{1}{2} \ln{\left( \frac{(v_a \cdot n)^2}{\vert n \vert^2}\frac{(v_b \cdot n)^2}{\vert n \vert^2}\right)} + i \pi - 1  \right], \\
 \omega^{a1} &=& S_{a1} \frac{\alpha_s}{\pi \epsilon} \left[ - \frac{1}{2} \ln{\left( \frac{(v_a \cdot v_1 )^2 s}{2 \mgl^2}\right)} + L_1 +  \frac{1}{2} \ln{\left( \frac{(v_a \cdot n)^2}{\vert n \vert^2}\right)} - 1  \right], \\
 \omega^{b1} &=& S_{b1} \frac{\alpha_s}{\pi \epsilon} \left[ - \frac{1}{2} \ln{\left( \frac{(v_b \cdot v_1 )^2 s}{2 \mgl^2}\right)} + L_1 +  \frac{1}{2} \ln{\left( \frac{(v_b \cdot n)^2}{\vert n \vert^2}\right)} - 1  \right], \\
 \omega^{11} &=& S_{11} \frac{\alpha_s}{\pi \epsilon} \left[ 2 L_1 - 2 \right].
\eea
Here, we have combined the signs in $S_{kl} = \Delta_k \Delta_l \delta_k \delta_l$, so that
$S_{ab} = 1$, $S_{a1} = 1$, $S_{b1} = -1$, and $S_{11} = -1$. The double poles in $\eps$ in the first
three integrals involving at least one massless particle have canceled among themselves.
As one can easily see, the scalar products are
\bea
 v_a \cdot v_b &=& \frac{2 p_a \cdot p_b}{s} = 1\,,\\
 v_a \cdot v_1 &=& \frac{2 p_a \cdot p_1}{s} = \frac{\mgl^2 - t}{s}\,,\\
 v_b \cdot v_1 &=& \frac{2 p_b \cdot p_1}{s} = \frac{\mgl^2 - u}{s} \,.
\eea
The function $L_k=[L_k(+n)+L_k(-n)]/2$ depends in a rather complicated way on the gauge vector $n$
\cite{Kidonakis:1997gm,Vincent}. However, all gauge-dependent terms disappear after the inclusion
of the self-energies of the two incoming Wilson lines.

Combining colour factors, signs, soft integrals and simplifying the result leads to
\beq
 {\bar\Gamma}_{q\bar{q} \rightarrow \tilde{g} \tilde{\chi}} =
 \frac{\alpha_s}{2\pi} C_A \left[\ln{2} + i \pi - 1
 + \ln{\left( \frac{\mgl^2 - t} {\sqrt{2}\mgl\sqrt{s}}\right)}
 + \ln{\left( \frac{\mgl^2 - u }{\sqrt{2} \mgl \sqrt{s}}\right)}\right]
 = \frac{\alpha_s}{2 \pi} C_A \left( T_{a1} + T_{b1} \right)
\eeq
with
\bea
 T_{a1} = \ln{\left( \frac{\mgl^2 - t}{\mgl \sqrt{s}}\right)} + \frac{i \pi - 1}{2}&\ ,\ &
 T_{b1} = \ln{\left( \frac{\mgl^2 - u}{\mgl \sqrt{s}}\right)} + \frac{i \pi - 1}{2}.
\eea
Due to the LSZ reduction formula, only half of the self-energy contribution
$\omega^{11}$ has been taken into account. All terms proportional to $C_F$ have vanished,
so that only terms proportional to $C_A$ remain. In the massless limit and before subtracting
the initial-state self-energies, one recovers the well-known result for associated gluon-photon
production, {\it i.e.}~the $C_A$-term in Eq.~(2.26) of Ref.~\cite{Kidonakis:1999hq}. Our modified
soft anomalous dimension can also be compared to the one obtained for associated top-quark and $W$-boson
production in Eq.~(3.8) of Ref.~\cite{Kidonakis:2006bu} after adjustments of the colour factors.
The result in Eq.~(3.1) of Ref.~\cite{Kidonakis:2010ux} is slightly different, since Feynman
gauge and not axial gauge was used there.

The final result for the soft wide-angle emission function in associated gluino-gaugino production is
therefore
\beq
 \label{eq:2.38}
 D_{q\bar{q} \rightarrow \tilde{g} \tilde{\chi}} = \operatorname{Re}
 \left[C_A \left( T_{a1} + T_{b1} \right)\right]\,.
\eeq
At the production threshold, where the final-state particle velocities vanish and
\beq
 \beta = \sqrt{1 - \frac{(\mgl + \mga)^2}{s}} \rightarrow 0\,,
\eeq
we find
\beq
 D_{q\bar{q} \rightarrow \tilde{g} \tilde{\chi}} = - C_A\,,
\eeq
in accordance with Ref.\ \cite{Beenakker:2009ha}.

\section{Lists of total cross sections}
\label{sec:c}

In Tabs.\ \ref{tab:c1} and \ref{tab:c2} we list the total cross sections
for the associated production of a second-lightest neutralino and a gluino
at the LHC with a centre-of-mass energy of 13 TeV in tabular form.
In Tab.\ \ref{tab:c1}, these cross sections are presented as a function
of the gluino mass, while in Tab.\ \ref{tab:c2} they are presented as a 
function of the wino mass parameter $M_2$. These tables thus correspond
to Figs.\ \ref{fig:06} and \ref{fig:08}. They also include listings of the
respective scale and PDF uncertainties.

\begin{table}[htbp] 
\centering 
\begin{tabular}{|Sc|Sc Sc Sc|}
\toprule
$\mgl \, (\mathrm{GeV})$ & $\mathrm{LO}^{+\mathrm{scale}}_{-\mathrm{scale}} \, (\mathrm{fb})$ & $\mathrm{NLO}^{+\mathrm{scale} + \mathrm{PDF}}_{-\mathrm{scale} - \mathrm{PDF}} \, (\mathrm{fb})$ & $\mathrm{NLL}^{+\mathrm{scale}}_{-\mathrm{scale}} \, (\mathrm{fb})$ \\ 
\hline 
$500$ & $2.211^{+18.7\%}_{-14.8\%}$ & $2.827^{+11.0\% \, +6.2\%}_{-9.5\%\, -8.1\%}$ & $2.751^{+8.0\%}_{-5.8\%}$ \\
$600$ & $1.722^{+19.1\%}_{-15.1\%}$ & $2.044^{+9.5\% \, +8.7\%}_{-8.7\%\, -6.8\%}$ & $2.016^{+7.0\%}_{-5.2\%}$ \\
$700$ & $1.350^{+19.5\%}_{-15.3\%}$ & $1.543^{+8.2\% \, +9.1\%}_{-8.4\%\, -7.3\%}$ & $1.532^{+5.7\%}_{-4.8\%}$ \\
$800$ & $1.065^{+19.9\%}_{-15.5\%}$ & $1.185^{+7.4\% \, +9.1\%}_{-7.9\%\, -8.8\%}$ & $1.178^{+4.7\%}_{-4.2\%}$ \\
$900$ & $0.844^{+20.2\%}_{-15.8\%}$ & $0.920^{+7.4\% \, +9.3\%}_{-7.8\%\, -9.4\%}$ & $0.917^{+4.6\%}_{-4.0\%}$ \\
$1000$ & $0.671^{+20.5\%}_{-16.0\%}$ & $0.719^{+6.7\% \, +9.6\%}_{-8.3\%\, -9.5\%}$ & $0.721^{+3.8\%}_{-4.3\%}$ \\
$1100$ & $0.536^{+20.8\%}_{-16.2\%}$ & $0.563^{+6.8\% \, +10.7\%}_{-7.8\%\, -9.8\%}$ & $0.566^{+3.6\%}_{-3.6\%}$ \\
$1200$ & $0.429^{+21.1\%}_{-16.3\%}$ & $0.444^{+6.7\% \, +10.1\%}_{-7.6\%\, -11.3\%}$ & $0.447^{+3.2\%}_{-3.4\%}$ \\
$1300$ & $0.344^{+21.4\%}_{-16.5\%}$ & $0.350^{+6.6\% \, +11.4\%}_{-7.3\%\, -11.1\%}$ & $0.353^{+3.0\%}_{-2.9\%}$ \\
$1400$ & $0.277^{+21.6\%}_{-16.7\%}$ & $0.277^{+6.2\% \, +11.6\%}_{-7.6\%\, -12.2\%}$ & $0.280^{+2.5\%}_{-3.1\%}$ \\
$1500$ & $0.224^{+21.9\%}_{-16.9\%}$ & $0.220^{+6.7\% \, +11.7\%}_{-7.3\%\, -13.3\%}$ & $0.223^{+2.9\%}_{-2.7\%}$ \\
$1600$ & $0.180^{+22.2\%}_{-17.0\%}$ & $0.174^{+6.4\% \, +13.4\%}_{-6.8\%\, -12.9\%}$ & $0.177^{+2.3\%}_{-2.1\%}$ \\
$1700$ & $0.146^{+22.4\%}_{-17.2\%}$ & $0.139^{+5.6\% \, +14.2\%}_{-7.1\%\, -13.6\%}$ & $0.141^{+1.5\%}_{-2.2\%}$ \\
$1800$ & $0.118^{+22.7\%}_{-17.4\%}$ & $0.110^{+5.4\% \, +15.0\%}_{-7.1\%\, -14.3\%}$ & $0.113^{+1.3\%}_{-2.1\%}$ \\
$1900$ & $0.096^{+22.9\%}_{-17.5\%}$ & $0.088^{+5.7\% \, +15.6\%}_{-7.1\%\, -15.1\%}$ & $0.090^{+1.3\%}_{-2.0\%}$ \\
$2000$ & $0.078^{+23.2\%}_{-17.7\%}$ & $0.071^{+5.4\% \, +16.4\%}_{-6.9\%\, -16.1\%}$ & $0.073^{+1.0\%}_{-1.7\%}$ \\
$2100$ & $0.063^{+23.4\%}_{-17.8\%}$ & $0.057^{+5.4\% \, +17.9\%}_{-7.3\%\, -16.2\%}$ & $0.059^{+0.9\%}_{-1.9\%}$ \\
$2200$ & $0.051^{+23.7\%}_{-18.0\%}$ & $0.046^{+5.4\% \, +18.2\%}_{-7.3\%\, -17.3\%}$ & $0.048^{+0.7\%}_{-1.8\%}$ \\
$2300$ & $0.042^{+23.9\%}_{-18.1\%}$ & $0.037^{+5.6\% \, +19.3\%}_{-7.2\%\, -18.0\%}$ & $0.038^{+1.0\%}_{-1.6\%}$ \\
$2400$ & $0.034^{+24.1\%}_{-18.3\%}$ & $0.030^{+5.6\% \, +20.3\%}_{-7.3\%\, -18.8\%}$ & $0.031^{+1.0\%}_{-1.7\%}$ \\
$2500$ & $0.028^{+24.4\%}_{-18.4\%}$ & $0.024^{+5.7\% \, +21.3\%}_{-7.6\%\, -19.7\%}$ & $0.025^{+0.8\%}_{-1.8\%}$ \\
$2600$ & $0.022^{+24.6\%}_{-18.5\%}$ & $0.019^{+5.9\% \, +22.8\%}_{-7.6\%\, -20.4\%}$ & $0.020^{+0.9\%}_{-1.7\%}$ \\
$2700$ & $0.018^{+24.8\%}_{-18.7\%}$ & $0.016^{+6.1\% \, +23.4\%}_{-7.8\%\, -21.7\%}$ & $0.017^{+0.8\%}_{-1.8\%}$ \\
$2800$ & $0.015^{+25.1\%}_{-18.9\%}$ & $0.013^{+6.3\% \, +24.9\%}_{-8.1\%\, -22.4\%}$ & $0.013^{+0.7\%}_{-1.8\%}$ \\
$2900$ & $0.012^{+25.3\%}_{-19.0\%}$ & $0.010^{+6.4\% \, +26.3\%}_{-8.2\%\, -23.3\%}$ & $0.011^{+0.7\%}_{-1.8\%}$ \\
$3000$ & $0.010^{+25.5\%}_{-19.2\%}$ & $0.008^{+6.6\% \, +27.7\%}_{-8.3\%\, -24.3\%}$ & $0.009^{+0.6\%}_{-1.8\%}$ \\
\bottomrule
\end{tabular}
\caption{\label{tab:c1}Total cross sections for $p\,p \rightarrow \tilde{\chi}_2^0 \, \tilde{g}$ at $\sqrt{S} = 13\, \mathrm{TeV}$ as a function of the gluino mass $\mgl$ using CT14NLO PDFs.}
\end{table}

\begin{table}[htbp] 
\centering 
\begin{tabular}{|Sc|Sc Sc Sc|}
\toprule
$M_2 = 2 \, M_1 \, (\mathrm{GeV})$ & $\mathrm{LO}^{+\mathrm{scale}}_{-\mathrm{scale}} \, (\mathrm{fb})$ & $\mathrm{NLO}^{+\mathrm{scale} + \mathrm{PDF}}_{-\mathrm{scale} - \mathrm{PDF}} \, (\mathrm{fb})$ & $\mathrm{NLL}^{+\mathrm{scale}}_{-\mathrm{scale}} \, (\mathrm{fb})$ \\ 
\hline 
$100$ & $0.309^{+21.9\%}_{-16.9\%}$ & $0.272^{+2.7\% \, +13.0\%}_{-5.4\%\, -13.2\%}$ & $0.277^{+0.6\%}_{-0.9\%}$ \\
$150$ & $0.271^{+22.0\%}_{-17.0\%}$ & $0.239^{+3.0\% \, +13.9\%}_{-5.6\%\, -13.3\%}$ & $0.245^{+0.6\%}_{-1.1\%}$ \\
$200$ & $0.238^{+22.2\%}_{-17.0\%}$ & $0.211^{+3.4\% \, +13.7\%}_{-5.6\%\, -14.0\%}$ & $0.215^{+0.7\%}_{-1.0\%}$ \\
$250$ & $0.208^{+22.3\%}_{-17.1\%}$ & $0.185^{+3.5\% \, +14.4\%}_{-5.9\%\, -13.9\%}$ & $0.189^{+0.6\%}_{-1.1\%}$ \\
$300$ & $0.182^{+22.4\%}_{-17.2\%}$ & $0.162^{+3.8\% \, +14.9\%}_{-6.2\%\, -13.9\%}$ & $0.166^{+0.7\%}_{-1.4\%}$ \\
$350$ & $0.159^{+22.6\%}_{-17.3\%}$ & $0.142^{+4.3\% \, +14.6\%}_{-6.1\%\, -14.8\%}$ & $0.145^{+0.8\%}_{-1.2\%}$ \\
$400$ & $0.139^{+22.7\%}_{-17.4\%}$ & $0.124^{+4.3\% \, +15.2\%}_{-6.3\%\, -14.9\%}$ & $0.128^{+0.8\%}_{-1.3\%}$ \\
$450$ & $0.121^{+22.8\%}_{-17.5\%}$ & $0.109^{+4.6\% \, +15.7\%}_{-6.6\%\, -15.3\%}$ & $0.112^{+0.8\%}_{-1.6\%}$ \\
$500$ & $0.105^{+23.0\%}_{-17.5\%}$ & $0.095^{+4.7\% \, +16.2\%}_{-6.8\%\, -15.2\%}$ & $0.098^{+0.7\%}_{-1.7\%}$ \\
$550$ & $0.092^{+23.1\%}_{-17.6\%}$ & $0.083^{+4.9\% \, +16.5\%}_{-6.8\%\, -15.4\%}$ & $0.085^{+0.8\%}_{-1.7\%}$ \\
$600$ & $0.078^{+23.2\%}_{-17.7\%}$ & $0.071^{+5.1\% \, +16.9\%}_{-7.0\%\, -15.7\%}$ & $0.073^{+0.9\%}_{-1.8\%}$ \\
$650$ & $0.065^{+23.3\%}_{-17.8\%}$ & $0.059^{+5.5\% \, +17.0\%}_{-7.2\%\, -16.4\%}$ & $0.061^{+0.9\%}_{-1.9\%}$ \\
$700$ & $0.051^{+23.4\%}_{-17.8\%}$ & $0.046^{+6.0\% \, +18.4\%}_{-7.1\%\, -15.7\%}$ & $0.048^{+1.4\%}_{-1.8\%}$ \\
$750$ & $0.036^{+23.4\%}_{-17.8\%}$ & $0.032^{+5.8\% \, +18.3\%}_{-7.4\%\, -16.1\%}$ & $0.033^{+1.0\%}_{-1.9\%}$ \\
$800$ & $0.022^{+23.4\%}_{-17.8\%}$ & $0.020^{+5.8\% \, +17.9\%}_{-7.4\%\, -16.7\%}$ & $0.020^{+1.0\%}_{-1.9\%}$ \\
$850$ & $0.013^{+23.3\%}_{-17.8\%}$ & $0.012^{+6.0\% \, +18.8\%}_{-7.5\%\, -16.1\%}$ & $0.012^{+1.1\%}_{-1.9\%}$ \\
$900$ & $0.008^{+23.1\%}_{-17.7\%}$ & $0.007^{+5.8\% \, +18.2\%}_{-7.4\%\, -16.8\%}$ & $0.008^{+0.8\%}_{-1.8\%}$ \\
$950$ & $0.005^{+22.9\%}_{-17.5\%}$ & $0.005^{+5.9\% \, +17.9\%}_{-7.2\%\, -17.1\%}$ & $0.005^{+0.8\%}_{-1.5\%}$ \\
$1000$ & $0.004^{+22.7\%}_{-17.4\%}$ & $0.004^{+5.9\% \, +18.8\%}_{-7.3\%\, -16.5\%}$ & $0.004^{+0.7\%}_{-1.5\%}$ \\
$1050$ & $0.004^{+22.5\%}_{-17.3\%}$ & $0.003^{+5.8\% \, +18.8\%}_{-7.1\%\, -16.5\%}$ & $0.003^{+0.5\%}_{-1.4\%}$ \\
$1100$ & $0.003^{+22.3\%}_{-17.2\%}$ & $0.003^{+6.0\% \, +18.9\%}_{-7.2\%\, -16.5\%}$ & $0.003^{+0.7\%}_{-1.4\%}$ \\
$1150$ & $0.003^{+22.3\%}_{-17.2\%}$ & $0.003^{+6.0\% \, +18.9\%}_{-7.2\%\, -16.6\%}$ & $0.003^{+0.7\%}_{-1.5\%}$ \\
$1200$ & $0.003^{+22.3\%}_{-17.2\%}$ & $0.003^{+6.1\% \, +18.9\%}_{-7.2\%\, -16.2\%}$ & $0.003^{+0.8\%}_{-1.4\%}$ \\
$1250$ & $0.003^{+22.4\%}_{-17.2\%}$ & $0.003^{+6.2\% \, +18.9\%}_{-7.4\%\, -16.1\%}$ & $0.003^{+0.8\%}_{-1.6\%}$ \\
$1300$ & $0.004^{+22.6\%}_{-17.3\%}$ & $0.004^{+6.3\% \, +18.7\%}_{-7.6\%\, -16.0\%}$ & $0.004^{+0.7\%}_{-1.8\%}$ \\
\bottomrule
\end{tabular}
\caption{\label{tab:c2}Total cross sections for $p\,p \rightarrow \tilde{\chi}_2^0 \, \tilde{g}$ at $\sqrt{S} = 13\, \mathrm{TeV}$ as a function of the wino mass parameter $M_2$ using CT14NLO PDFs.}
\end{table}

\bibliographystyle{JHEP}
\bibliography{paper}

\end{document}